\documentclass[preprint2]{aastex}

\def\mm{M\'endez }

\begin{document}

\title{An Atlas of Burst Oscillations and Spectral Properties in 4U 1728--34}

\author{Steve van Straaten\altaffilmark{1}, Michiel van der Klis\altaffilmark{1}, 
Erik Kuulkers\altaffilmark{2,3}, Mariano \mm\altaffilmark{1,4}}

\email{straaten@astro.uva.nl}
\altaffiltext{1}{Astronomical Institute, ``Anton Pannekoek'',
University of Amsterdam, and Center for High Energy Astrophysics, 
Kruislaan 403, 1098 SJ Amsterdam, The Netherlands.}
\altaffiltext{2}{Space Research Organization Netherlands, Sorbonnelaan 2, 
3584 CA Utrecht, The Netherlands.}
\altaffiltext{3}{Astronomical Institute, Utrecht University, P.O. Box 80000, 
3507 TA Utrecht, The Netherlands.}
\altaffiltext{4}{Facultad de Ciencias Astron\'omicas y Geof\'{\i}sicas,
Universidad Nacional de La Plata, Paseo del Bosque S/N, 1900 La Plata,
Argentina.}


\begin{abstract}

We study correlations between the burst properties and the position 
in the color--color diagram of the low mass X-ray binary 4U 1728--34 
using 21 bursts found in two data sets obtained with the Rossi X-ray Timing 
Explorer. Using spectral fits we analyze the spectral evolution during the bursts and 
determine the burst peak flux, temperature, rise time and fluence.
Using dynamical power spectra we study the properties of the $\sim 363$ Hz burst 
oscillations. 

We find correlations between fluence, peak flux, the occurrence of radius expansion and 
the presence of burst oscillations, and the position of the source in the color--color diagram. 
The radius expansion bursts with and without burst oscillations differ both with respect 
to where they occur in the color--color diagram and in how the radius expansion takes place.
We compare our results with those of a similar study by Muno et al. (2000) for 
KS 1731--26. Both KS 1731--260 and 4U 1728--34 show more or less the same behavior with respect 
to the dependence of the presence/absence of the 
burst oscillations on the position of the source in the color--color diagram, but the 
dependence of most of the spectral burst parameters on the position in the color--color 
diagram is clearly different. We find that the systematic absence in 4U 1728--34 of burst  
oscillations in the ``island'' state of the color--color diagram is not explained by the 
bursts involving mixed H--He burning, as may be the case in KS 1731--260 (Cumming \& 
Bildsten 2000).

\end{abstract}

\keywords{accretion, accretion disks --- binaries: close --- 
stars: individual (4U 1728--34) --- stars: neutron --- X--rays: bursts 
X--rays: stars}

\section{Introduction} 

Type I X--ray bursts (for a review, see, e.g., Lewin et al. 1993) in the low mass 
X--ray binary (LMXB) 4U 1728--34 were 
discovered with the SAS--3 observatory in 1976 (Lewin 1976; Hoffman et al. 
1976). An extensive study of the burst properties of 96 bursts was done by 
Basinska et al. (1984) with the Y-axis detectors on board SAS 3. Radius 
expansion bursts were reported by Basinska et al. (1984), Inoue (1988), Foster 
et al. (1986) and Day \& Tawara (1990). The source has been classified 
by Hasinger \& van der Klis (1989) as an atoll source based on timing and 
spectral observations with the EXOSAT satellite. The distance to the source is 
about 4 kpc (van Paradijs 1978; Foster et al. 1986; 
Di Salvo et al. 2000b). No optical counterpart has been identified (Liller 1977).

After the launch of the Rossi X--ray Timing Explorer (RXTE), burst oscillations 
on a millisecond time scale (i.e. at 363 Hz) were first discovered in 4U 1728--34 by Strohmayer 
et al. (1996a). Such oscillations have since been found in several other 
atoll sources, i.e. 4U 1636--53 
(Strohmayer et al. 1998a,b; Strohmayer 1999; Miller 1999, 2000; Zhang et al. 
1997) , 4U 1702--43 (Strohmayer \& Markwardt 1999; Markwardt et al. 1999), 
KS 1731--260 (Morgan \& Smith 1996; Smith et al. 1997; Muno et al. 2000), MXB 
1743--29 (Strohmayer et al. 1996b; Strohmayer et al. 1997), Aql X--1 (Zhang et 
al. 1998; Ford 1999; Yu et al. 1999), and marginally in the Rapid Burster (Fox \& 
Lewin 1999). The frequency during the burst usually increases to an asymptotic value which 
tends to be quite stable (see for an overview Strohmayer, Swank, \& Zhang 1998). This led to 
a simple model (Strohmayer et al. 1997) in which the oscillations arise due to 
a hot spot in a layer that due to the burst has expanded in radius (by 10--30 
m, this is not related to the expansion by several kilometers during classical radius expansion bursts)
and is therefore rotating slightly slower than the neutron star itself. As the 
layer recontracts the frequency increases towards 
the neutron star spin frequency. However, note that in 4U 1636--53 (Miller 
1999; Strohmayer 1999) and in KS 1731--260 (Muno et al. 2000) decreases in 
frequency have also been seen. For 4U 1728--34 Strohmayer et al. (1998b) found that for 3 bursts 
(bursts 4, 5 and 17 in this paper) occurring more than 1.5 yr apart, the 
observed asymptotic oscillation periods were all close to 2.7476 ms and within 0.2 
$\mu s$ of each other.

For several atoll sources correlations have been found between timing properties and the position 
of the source in the X--ray color--color diagram (e.g., Hasinger \& van der Klis 1989; 
Prins \& van der Klis 1997; \mm et al. 1999; van Straaten et al. 2000). A correlation 
between the position of 4U 1728--34 in the color--color diagram and several 
power spectral features is shown in \mm \& van der Klis (1999) and Di Salvo 
et al. (2000a). The most obvious explanation for these correlations is that both 
the timing properties and the position in the color--color diagram track the 
changing mass accretion rate (Hasinger \& van der Klis 1989), from low in the 
so--called island state, and increasing along the banana branch (for definitions see 
Hasinger \& van der Klis 1989). A link between 
the position in the color--color diagram and burst spectral properties was found 
for 4U 1636--53 (van der Klis et al. 1990) and 4U 1705--44 (van der Klis 1989).  

For KS 1731--260 Muno et al. (2000)  
connected the position in the color--color diagram with both the X-ray spectral 
properties during the bursts and the presence of burst oscillations. 
In the island state, KS 1731--260 showed longer decay time scales, larger fluences, 
no evidence for radius expansion and no burst oscillations. On the banana branch, 
the bursts were shorter, the fluences were smaller, 
and radius expansion episodes (where the emitting region expands by a factor of 2.5--5.2) 
and burst oscillations were present. 

In this paper we present an overview of the spectral and burst--oscillation properties of 
21 bursts in 4U 1728--34. We particularly investigate the relation between oscillation 
characteristics and radius--expansion characteristics of the bursts. We also connect our results 
to the position of the source in the color-color diagram when the bursts took place.

\section{Observations and Data Analysis}

In this analysis we use data obtained with RXTE's proportional counter 
array (PCA; for more instrument information see Zhang et al. 1993).
To prevent break down events the voltage of the PCA was lowered 
on three occasions leading to four periods of a different voltage, called gain epochs.
We study bursts occurring in two RXTE data sets. One set (observation ID 10073) 
took place during gain {\em epoch 1} (before 1996 March 21), and one (observation ID 
20083) during gain {\em epoch 3} (between 1996 April 15 and 1999 March 22). We find a 
total number of 21 bursts. In the data we analyzed all 5 detectors of the PCA were 
operational. All bursts are listed in Table \ref{tbl:obs}. The start times listed in 
Table \ref{tbl:obs} are the times that the rise in count rate, caused by the burst, 
triggered the Experiment Data System (EDS). 

\subsection{Spectral Analysis}

We use the Standard 2 mode, which has a 16--s time--resolution 
and 129 energy channels, to analyze the spectral properties of  
96 s of persistent emission just before each burst. To study 
the spectral properties during the bursts we use a burst trigger 
mode which has 64 energy channels and a time resolution of 8 ms. 
We determine a spectrum every 0.25 s. We correct each spectrum 
for background and dead--time using the methods supplied by the 
RXTE team. We fit the spectra between 2.8 and 20 keV. For the fits 
we implement a systematic error of 1 \%. We fix the hydrogen column 
density, $N_{\rm H}$, to the value of $2.5 \times 10^{22}$ atoms 
cm$^{-2}$ (Hoffman et al. 1976; Grindlay \& Hertz 1981; Foster et al. 
1986; Di Salvo et al. 2000b). 

The spectra of the persistent emission were fit with an exponentially 
cut--off power law defined as $N_{\rm \Gamma} E^{\rm -\Gamma} \exp^{\rm -E/kT}$, where 
$E$ is the photon energy, $\Gamma$ the power--law index, $N_{\rm \Gamma}$ 
is the cut--off power law normalization, and $kT$ is the cut--off energy. To this we add 
a Gaussian line with a fixed width ($\sigma = 0.1$ keV) and energy (6.5 keV) as observed 
previously (see Di Salvo et al. 2000b). This line could be due to iron 
emission. For the epoch 1 data eight spectra have a reduced chi squared, 
$\chi^2$/dof $< 1.6$ and four spectra (the four with the lowest spectral
index, all $\sim$ 1) have a $\chi^2$/dof 
around 2.2 with 49 degrees of freedom. The fits are generally better for the data from epoch 3 
($\chi^2$/dof $< 1.2$ with 42 degrees of freedom) even for spectra as hard or harder than seen 
in epoch 1, presumably due to better calibration in epoch 3.

\begin{figure}
\epsscale{0.7}
\plotone{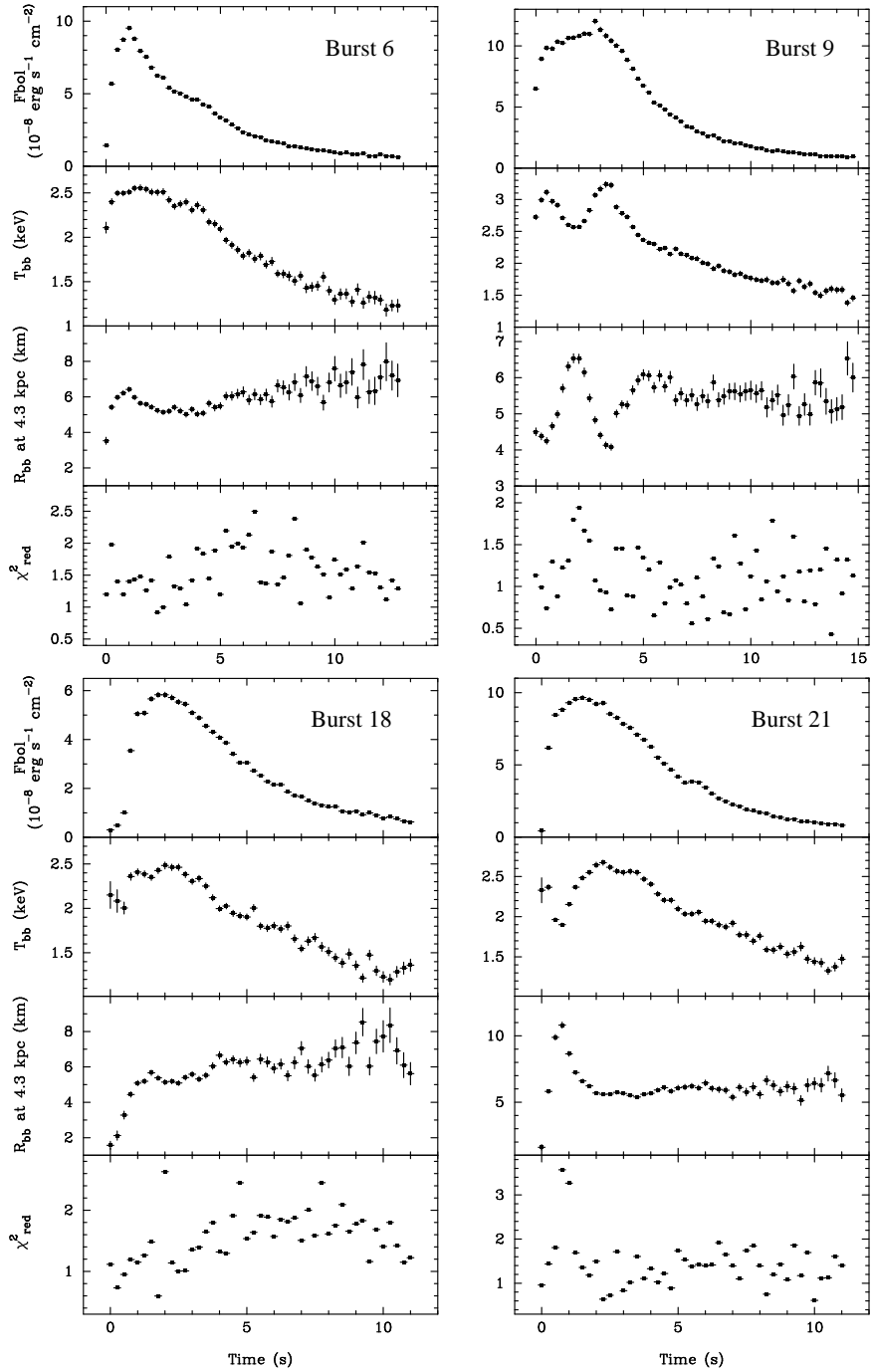}
\caption{Results of the spectral fitting during the burst for four 
representative bursts (bursts 6, 9, 18 and 21). 
Plotted are from top to bottom the bolometric blackbody flux, the blackbody 
temperature, the blackbody radius (at 4.3 kpc) and the reduced $\chi^2$.
Burst 21 shows classical radius expansion, burst 6 shows a hint for classical radius 
expansion (see \S 2.1), 
burst 9 shows radius expansion with unusual radius variations (see \S 2.1), and burst 18 shows 
no classical radius expansion.}
\label{fig:spectra}
\end{figure}

When we fit the burst spectra we model the total burst emission by a component equal to that 
of the persistent emission (which is fixed during the burst) plus a single blackbody component 
to describe the net burst emission. The bolometric blackbody flux, blackbody 
temperature, blackbody radius at 4.3 kpc (see Foster et al. 1986) and $\chi^2$/dof 
of four representative bursts are shown in Figure \ref{fig:spectra}. The fits generally have a $\chi^2$/dof below 
about 2 with 26 (epoch 1) or 23 (epoch 3) degrees of freedom, however sometimes it goes up 
to about 3. We note that in some of the 
bursts that show radius expansion (see below) the $\chi^2$/dof peaks during the increase/decrease 
phase of the radius (see e.g. bursts 21 and 9 in Figure \ref{fig:spectra}, this also occurs in bursts
1, 2, 7--11 and 20). It seems that the spectra 
deviate from simple Planckian functions when the radius varies rapidly (see also Lewin et al. 1993). 
We note that there 
are alternative methods (Sztajno et al. 1986) to handle the spectral analysis during bursts.
We picked the commonly used method described above as the aim of our paper  
is only to provide a homogeneous description of all burst properties for the purpose of 
correlating with the location in the color--color diagram (see below), and not to do detailed 
spectral modelling.

When a burst undergoes a radius expansion episode the luminosity $L$ stays constant and is 
equal to the Eddington luminosity, $L_{\rm edd}$ (modulo gravitational redshift effects with changing radius). 
Because $L \propto R^2 T_{\rm eff}^4$, with $R$ 
the radius of the photosphere and $T_{\rm eff}$ the effective temperature, when $R$ increases 
$T_{\rm eff}$ must decrease. Bursts 1, 2, 7--12, 20 and 21 clearly show this rise in blackbody 
radius and simultaneous dip in blackbody temperature. Bursts 4--6 have a peak flux comparable to that 
of these radius expansion bursts (see Table \ref{tbl:spec}), and therefore might be expected to show 
some evidence of radius expansion as well. Indeed, burst 6 shows a clear bump in radius accompanied by 
a slight dip in temperature (Fig. \ref{fig:spectra}) and bursts 4 and 5 only have one 0.25 s point were 
the radius peaked and the temperature dipped. Because the effects are marginal, hereafter we shall 
count those bursts as having no radius expansion. Bursts 13--18 show a gradual increase in blackbody 
radius, for the entire duration of the burst, but no simultaneous decrease in blackbody temperature. 
This is not classical radius expansion, but might either be a real increase in the X-ray burning area 
(see e.g. Kong et al. 2000) or it is a similar effect as that seen just after the radius 
contraction phase of bursts 8--12 (see below).

Bursts 8--12 show an unexpected spectral behaviour (see burst 9 in Fig. \ref{fig:spectra}). 
After the radius contraction phase 
the blackbody radius reaches a minimum value followed by an increase. This behaviour has 
previously been seen in bursts from several other sources (e.g. Hoffman et al. 1980; Tawara et 
al. 1984; van Paradijs et al. 1990; Kaptein et al. 2000), although we note that the durations 
of those bursts were all long ($> 60$ s). The behaviour could perhaps be due to a spectral hardening effect caused 
by electron scattering in the neutron star atmosphere, that 
depends mainly on the ratio $L/L_{\rm edd}$ (see for a discussion Kaptein et al. 2000). Kaptein 
et al. (2000) try to correct the temperature and radius for a burst in 1RXS J171824.2--402934 and 
find that when corrected the increase in radius after the minimum is substantially less (however, 
the effect does not disappear after correction).

From the spectral fits of the burst and the persistent emission we can determine the classical 
burst parameters such as the maximum flux (bolometric black body flux plus the persistent emmision integrated 
between 0.01--100 keV), $F_{\rm peak}$, the blackbody temperature at 
10 \% of the Eddington flux, $kT_{\rm 0.1}$, the rise time, $t_{\rm rise}$, and the fluence, 
$E_{\rm b}$. As an indicator for the temperature of a burst we use the temperature in the time bin closest 
to 10 \% of the Eddington flux in Figure \ref{fig:spectra}. The Eddington flux, $F_{\rm edd}$, is defined by
the highest peak flux from all the radius expansion bursts (i.e. that of burst 9: 
$12.6 \times 10^{-8}$ erg cm$^{-2}$ s$^{-1}$; at a distance of 4.3 
kpc and assuming that the burst emission is isotropic this leads to $L_{\rm edd}=2.79 \times 
10^{38}$ erg s$^{-1}$). To determine $E_{\rm b}$ we sum the flux 
from the start of the burst until the end of the available data. To compensate for the missing 
fluence at the end of the burst (since the burst flux decays exponentially it finishes at
infinite times), we fit the burst tail with 
an exponential function which we integrate from the end of the trigger file 
to infinity. We calculate the standard burst parameters (see e.g. Lewin et al. 1993) 
$\gamma$ ($=F_{\rm pers}/[F_{\rm peak}-F_{\rm pers}]$, where $F_{\rm pers}$ is the persistent flux just
before the burst), $\tau$ ($= E_b/[F_{\rm peak}-F_{\rm pers}]$) and a lower limit (because of South 
Atlantic Anomaly passages and earth occultations) for $\alpha$ ($=F_{\rm pers}/(E_b/\Delta 
t)$, where $\Delta t$ is the time since the previous burst). 
To determine the rise time, $t_{\rm rise}$ we use the PCA light curve over the full PCA energy 
band with a time resolution of 0.0312 s (see Figure \ref{fig:dynspec}). We define $t_{\rm rise}$ as the time for 
the burst to increase from 25 \% to 90 \% of the peak rate. All burst parameters are shown in 
Table \ref{tbl:spec}.

We can compare our values of $F_{\rm peak}$ with the peak bolometric flux found by Franco (2000, 
see also the note added to the manuscript) for the same set of bursts. Although they do not include the persistent 
emission in their listed values and extract spectra of 0.125 s instead of the 0.25 s we use, the two sets of 
results are in agreement.

\begin{figure}
\figurenum{3}
\epsscale{0.7}
\plotone{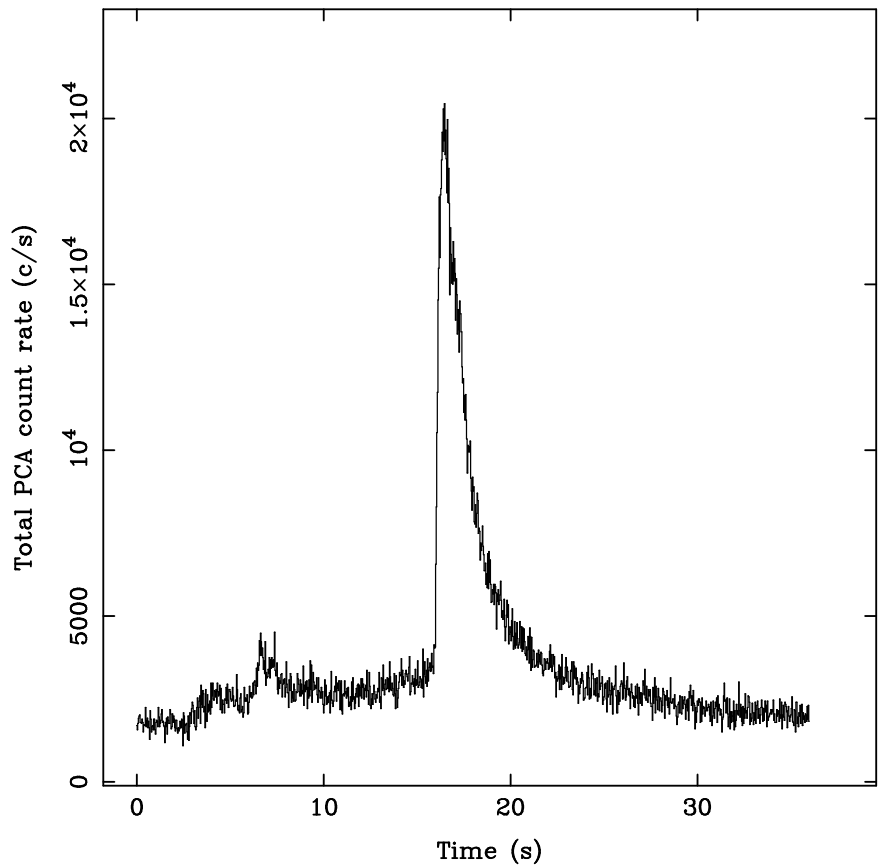}
\caption{The PCA lightcurve over the full PCA energy band of burst 19. Burst 19 is an anomalous burst 
that shows a rise in flux about 15 
s before the burst really goes off. In addition a small precursor occurs 9 s before the burst rise. 
The burst itself has a small fluence and is extremely short.}
\label{fig:burst19}
\end{figure}

The profile of burst number 19 deviates strongly from all the other bursts (see Fig. \ref{fig:burst19}). 
About 15 s before the burst really goes off the flux goes up and 9 s before the burst rise a small precursor 
occurs. The burst itself has a small fluence and is extremely short ($\tau = 2.0$ s) compared to 
the other bursts ($\tau > 4.3$ s). Including the pre-burst rise emission only increases the fluence 
by $\sim 30$ \%, so it remains low.

\subsection{Burst Oscillations}

To get a first indication of which bursts show oscillations we made Leahy normalized power 
spectra (Leahy et al. 1983) starting a few seconds before the burst with a Nyquist frequency 
of 2048 Hz and a frequency resolution of 0.25 Hz. We used a PCA data mode with a 125$\mu$s 
time resolution and no energy resolution (the PCA is sensitive in the 2--60 keV range). We 
determine if there is any power between 360 and 365 Hz (based on previous results by Strohmayer et al. 1998b) 
that is more significant than 3$\sigma$ (single trial). We find oscillations in 15 out of the 21 bursts. 
These preliminary detections were confirmed in our further analysis (see below).

We then examined the bursts with oscillations more closely. We made Leahy normalized dynamical 
power spectra starting about 4 s before the rise and extending to about 12 s after the rise. 
The dynamical power spectrum is composed of individual power spectra each covering a 4 s interval. 
These intervals overlap, each new interval starting 0.125 s later than the previous one, 
increasing the time resolution at the expense of introducing correlations. 
We initially looked for oscillations between 300 and 400 Hz but found no evidence for 
oscillations outside the 360--365 Hz interval. We show 
the dynamical power spectra between 360 and 365 Hz for the bursts that show oscillations in Figure \ref{fig:dynspec}. 
For each individual power spectrum there are 21 frequency bins between 360 and 365 Hz. 
Due to the overlaps, adjacent power spectra are correlated; after each 4 s the power spectra become 
uncorrelated. In a dynamical power spectrum with a total duration of 16 s there are four independent 
power spectra, which leads to a total of 84 ($21 \times 4$) independent powers. For clarity we plot 
only those powers above a value of 9.2, the 99\% confidence detection level for a single trial. 
In each plot about one independent power can be expected to exceed this level due to only Poisson noise 
fluctuations. Such a power fluctuation can show up in several adjacent power 
spectra. In each plot we include the PCA light curve over the full PCA energy 
band. The light curve and power spectra are aligned such that each point in the dynamical power spectrum 
represents a power spectrum taken over 2 s before and 2 s after the corresponding point in the light curve.

For the four strongest frequency tracks (4, 5, 14 and 17) we list the frequency shift and frequency 
endpoint of the tracks seen in Figure \ref{fig:dynspec} in Table \ref{tbl:freq_shft}. We note that these values 
might be influenced by noise powers (on average one per plot, see above) accidentally situated near the 
beginning or end of a frequency track.

Because the frequency of the oscillation drifts during the 4 s interval for which each power 
spectrum is computed, the power spreads over a number of frequency bins. To compute 
the total power of the oscillation for each power spectrum we first find the highest power, $P_{\rm max}$, 
between 360 and 365 Hz. We then determine the Poisson level, $P_{\rm pois}$, by averaging 
between 1000 and 1500 Hz and add up all power ($P_{\rm tot}$) above the Poisson level in an interval 
of 4 Hz around the frequency bin containing $P_{\rm max}$. As the oscillation signal is deterministic, 
the error in the power of each frequency bin is equal to 
$P_{\rm pois}$ so the error in the total power is obtained by quadratically adding these powers over the 4 
Hz interval. From the resulting total oscillation power we calculate the rms fractional amplitude plotted 
in the bottom frames of Figure \ref{fig:dynspec}. We only plot the rms fractional amplitude when the corresponding $P_{\rm max}$ 
is larger than 9.2 and the corresponding $P_{\rm tot}$ differs more then 1 $\sigma$ from 0. When this is 
not the case we calculate an upper limit in the same way as for the bursts that show no oscillations 
(see below).

For the bursts that show no significant oscillations we perform an identical dynamic power spectral 
analysis as above. No significant oscillations are present. We add up all power above the Poisson level between 360 and 
365 Hz, $P_{\rm 360-365}$, calculate the $1 \sigma$ error in this power in the same way as for $P_{\rm tot}$ 
(see above) and use this to obtain a 97.5 \% confidence upper limit on $P_{\rm tot}$ ($P_{\rm u.l.}$ = 
$P_{\rm 360-365}$ + 2 $\times$ error). 

\section{Results}

\begin{figure}
\figurenum{4}
\epsscale{0.9}
\plotone{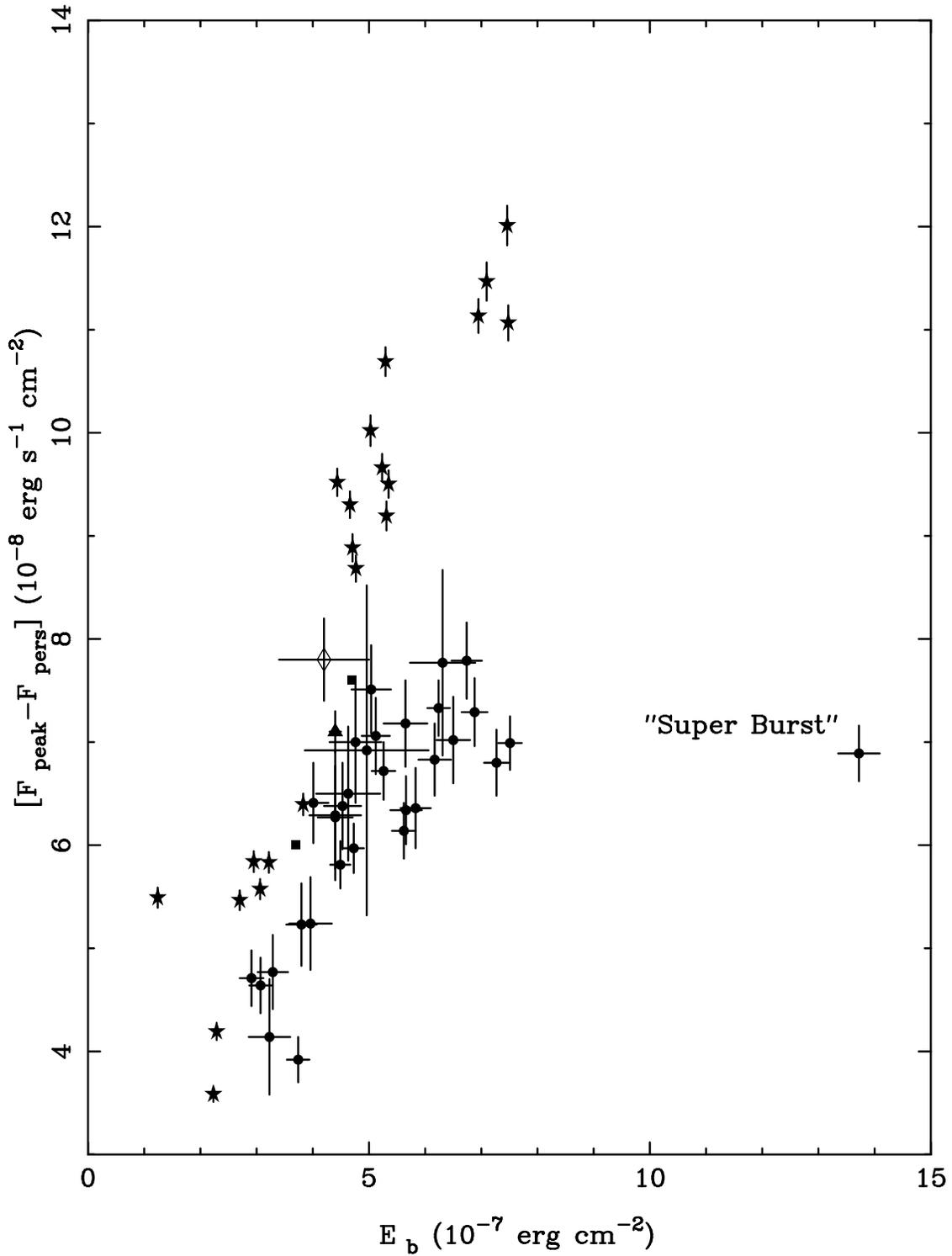}
\caption{$E_{\rm b}$ versus $F_{\rm peak}-F_{\rm pers}$ for 4U 1728--34. Filled circles are from Basinska et 
al. (1984), filled squares are from Day \& Tawara (1990), the filled triangle is from Hoffman (1979), 
the open triangle is from Kaminker et al. (1989). Filled stars are from this study. We have also marked 
the ``super burst'' from Basinska et al. (1984).}
\label{fig:basinska}
\end{figure}

Previous studies of spectral burst parameters in 4U 1728--34 have been made
with HEAO 1 (Hoffman et al. 1979), Einstein (Grindlay \& Hertz 1981), EXOSAT 
(Foster et al. 1986), Astron (Kaminker et al. 1989), and Ginga (Day \& 
Tawara 1990). The most extensive study was done with SAS 3 by Basinska et 
al. (1984) and included 96 burst detected in $\sim$48 days of observations 
between 1976 March and 1979 March. In Figure \ref{fig:basinska} we plot our values of $E_{\rm b}$ 
versus $F_{\rm peak}-F_{\rm pers}$ together with those of Basinska et al. (1984). We 
have also included two points obtained with Ginga (Day \& Tawara 1990), 
one with HEAO 1 (Hoffman et al. 1979), and one with Astron (Kaminker et al. 1989). Basinska et 
al. (1984) observed a strong, approximately linear, correlation between $E_{\rm b}$ 
and $F_{\rm peak}-F_{\rm pers}$ at low $E_{\rm b}$ levels; at higher $E_{\rm b}$ values there was 
a turnover at $7 \times 10^{-8}$ erg cm$^{-2}$ s$^{-1}$ and they suggested that 
the source had reached a critical luminosity. In our sample the strong 
correlation extends to values for $F_{\rm peak}-F_{\rm pers}$ up to $12 \times 
10^{-8}$ erg cm$^{-2}$ s$^{-1}$. We do not observe a turnover and $F_{\rm peak}-F_{\rm pers}$ 
clearly exceeds $7.5 \times 10^{-8}$ erg cm$^{-2}$ s$^{-1}$. We note that the measured 
values of $F_{\rm peak}-F_{\rm pers}$ depend on the size of the used time bins. However, this 
cannot account for the much larger $F_{\rm peak}-F_{\rm pers}$ we observe.
 
We find $\tau$ 
ranging between 4.7 and 6.8 s, with the exception of burst 19, which has $\tau = 
2.3$ s. Excluding the so called "Super Burst" (Basinska et al. 1994, see Figure 
\ref{fig:basinska}) which has $\tau = 19.9$ s Basinska et al. (1984) found 
$\tau$ to range between 5.1 and 12.3 s, and that the rise times were, with one 
exception, all short ($< 5$ s); this is consistent with our results.

\begin{figure}
\figurenum{5}
\epsscale{1.1}
\plottwo{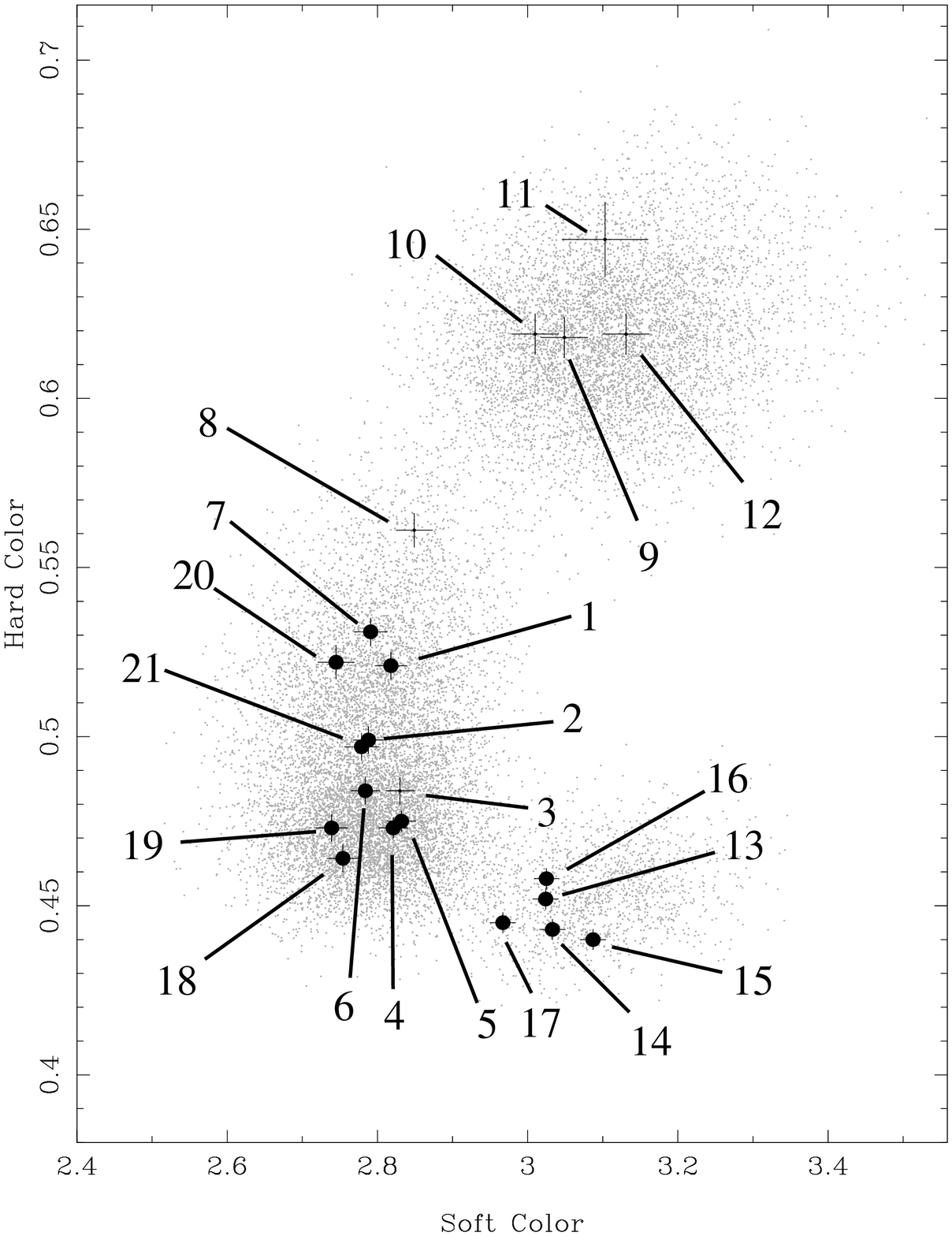}{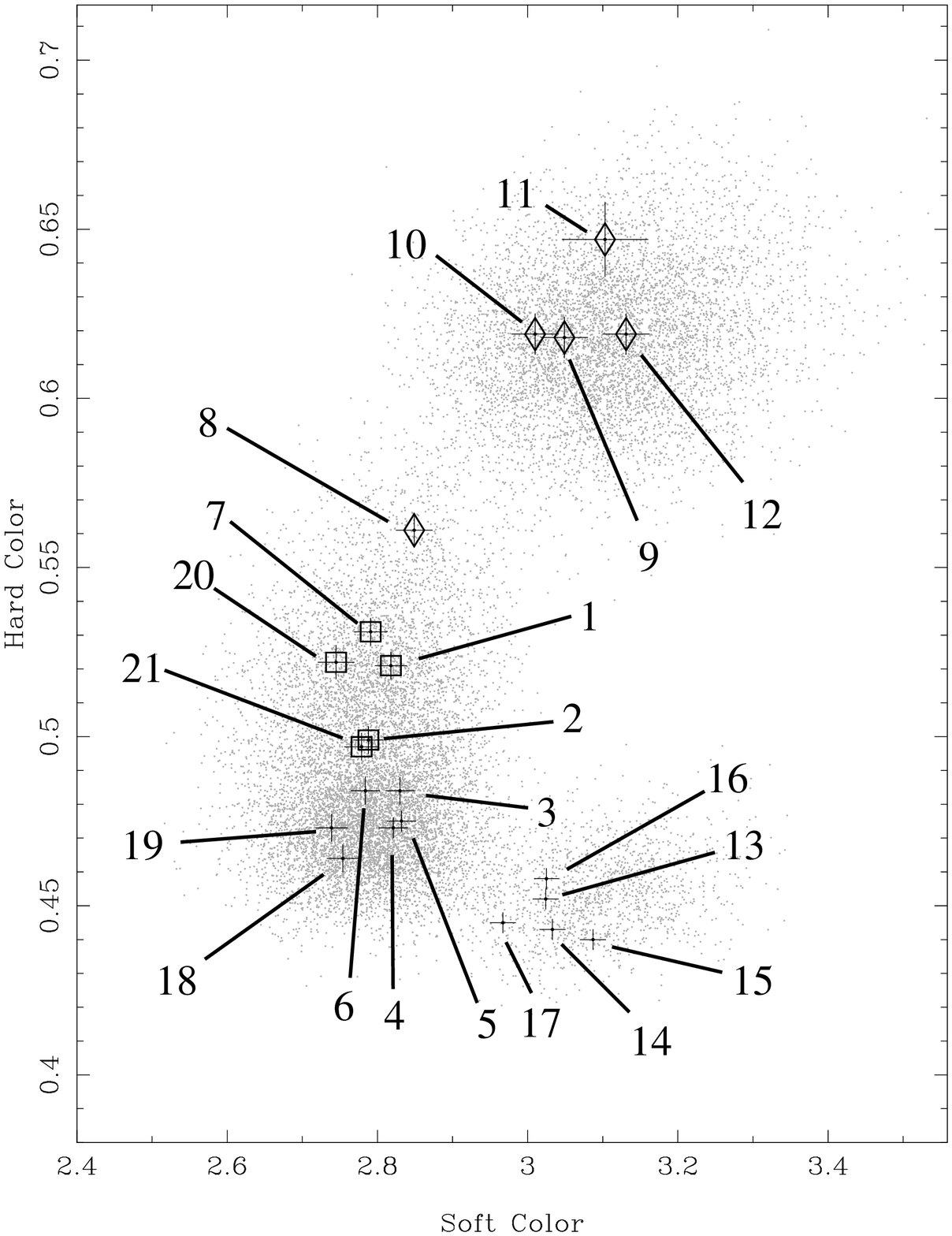}
\caption{
Color-color diagram of 4U 1728--34 (from Di Salvo et al. 2000a). 
The hard color is defined as the count rate in the energy band 9.7$-$16.0 keV 
divided by the rate in the energy band 6.4$-$9.7 keV and the soft color is defined 
as the count rate in the energy band 3.5$-$6.4 keV divided by the rate in the energy 
band 2.0$-$3.5 keV. We have indicated the position of the source in
this diagram prior to each of the 21 bursts. 
In the left--hand panel bursts with oscillations are marked with a filled 
circle, whereas bursts with no oscillations are marked with a plus symbol. 
In the right--hand panel bursts with radius expansion are marked with open symbols, 
the bursts with no radius expansion are marked with a plus symbol. Diamonds denote bursts
with the unusual radius variations described in \S 2.1, squares classical radius expansion 
bursts.}
\label{fig:cd}
\end{figure}

We determine the positions of the bursts in the color--color diagram produced 
by Di Salvo et al. (2000a) by taking the weighted average in hard and soft color 
for ten 16--s points from Standard 2 data prior to the burst. The hard color in Di Salvo 
et al. (2000a) is defined as the count rate in the energy band 9.7$-$16.0 keV 
divided by the rate in the energy band 6.4$-$9.7 keV and the soft color
as that in the energy band 3.5$-$6.4 keV divided by the rate in the energy 
band 2.0$-$3.5 keV. In burst 11, due to Earth occultation, only four 16 s points 
were available prior to the burst, leading to larger errors in the colors. The 
position of each burst in the color--color diagram is shown in Figure \ref{fig:cd}. 

For each burst we find the rank number as used by Di Salvo et al. (2000a) to mark the position 
in the color--color diagram. The soft color, hard color, rank number as well as the count 
rate in the 2.0--16.0 keV band are listed in Table \ref{tbl:obs}.
To study the correlation between the various burst spectral properties and position in the color--color 
diagram we plot several parameters versus their rank number in Figure \ref{fig:rn_vs_sp}. 
$E_{\rm b}$ and $F_{\rm peak}$ (and thus $\gamma$) clearly correlate with the position 
in the color color diagram, both decreasing with rank number (while $\gamma$ increases). 
Although $kT_{\rm 0.1}$ and $\tau$ change only 
marginally they also appear to decrease with rank number. 

In the right--hand panel of Figure \ref{fig:cd}, we marked the bursts with radius expansion with open 
symbols. The bursts with radius expansion that show the unusual radius variations described in \S 2.1 are 
marked with open diamonds. All the bursts in the island state, some of the bursts on the lower banana, 
and none of the bursts on the upper banana show radius expansion. In the left--hand panel of 
Figure \ref{fig:cd}, the bursts with oscillations are marked with filled circles, 
the ones without with pluses. The 5 bursts with rank numbers below 7 (island state) 
show no oscillations. These are precisely bursts 8--12, which exhibited the unusual radius variations 
(see \S 2.1). The bursts with rank numbers greater than 9 (the edge of the island state and 
the banana branch) all show oscillations (except the weak burst 3 at rank number 14, whose 
oscillation upper limit is consistent with the oscillation strength seen in the other bursts). 
These oscillating bursts therefore include all bursts with radius expansion that had no unusual 
radius variations, as well as all bursts (except burst 3) without radius expansion.

\begin{figure}
\figurenum{7}
\epsscale{0.9}
\plotone{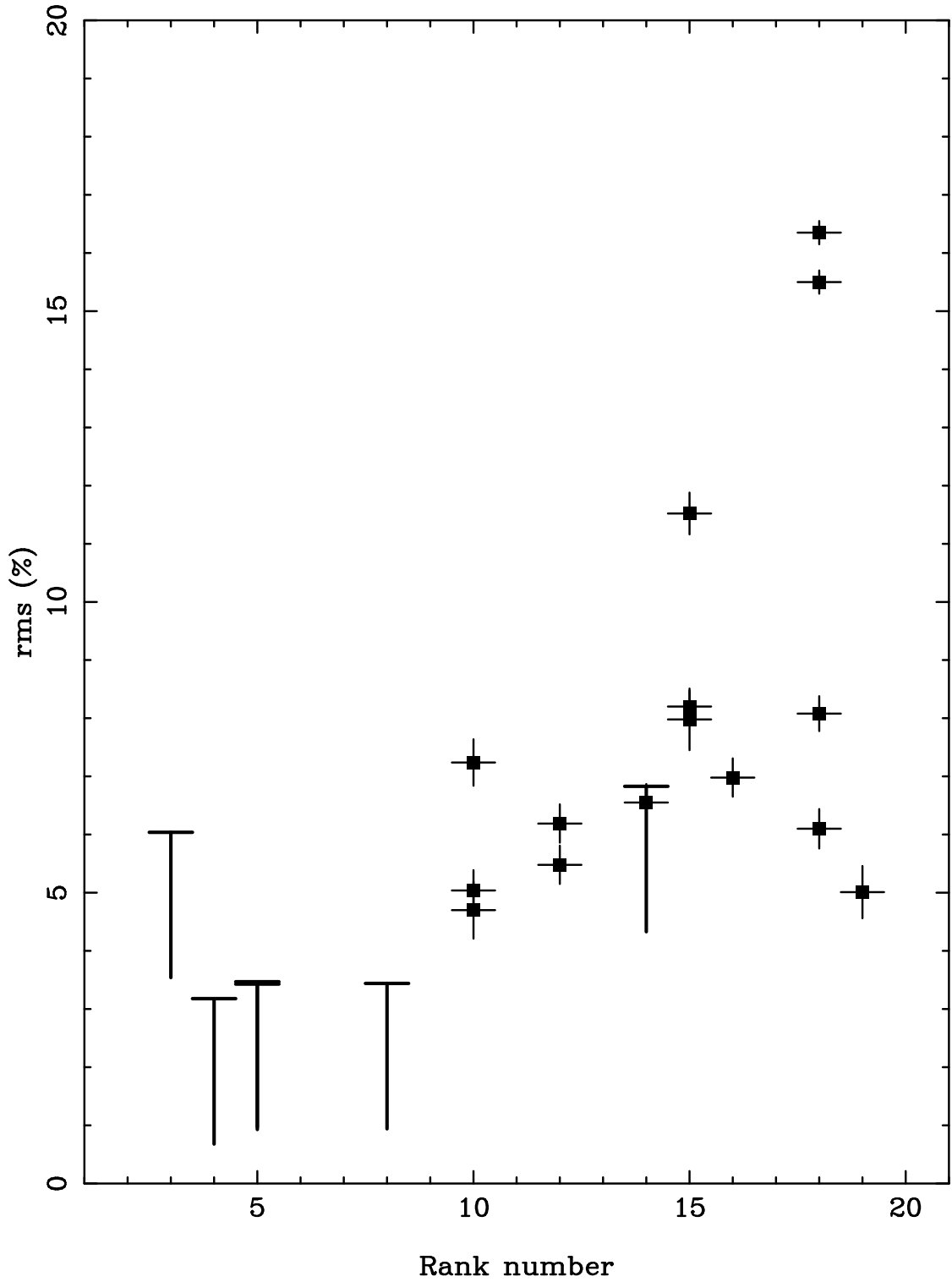}
\caption{The rms fractional amplitude of the burst oscillations during each burst plotted 
versus rank number.}
\label{fig:rms_vs_rn}
\end{figure}

Upper limits indicate that oscillations in the island state, if any, really have a lower rms 
than those on the banana branch and are not just harder to detect due to lower count rates there. 
To demonstrate this we find the rms fractional amplitude 
corresponding to the largest $P_{\rm tot}$ or $P_{\rm u.l.}$ in each burst. From this we exclude the first 
4 s of Figure \ref{fig:dynspec} where the resulting rms fractional amplitudes are systematically high because 
the summed count rate over 4 s contains a fraction of counts from the persistent emission.
The peak rms fractional amplitudes and upper limits are plotted in Figure \ref{fig:rms_vs_rn}. The higher peak 
rms fractional amplitudes tend to occur at higher rank numbers. 

The dynamical power spectra of the bursts with oscillations (Fig. \ref{fig:dynspec}) show several different types of 
frequency tracks. 
A remarkable difference can be seen between the frequency tracks of the bursts with oscillations 
{\it with} radius expansion (1, 2, 7, 20 and 21), and those without (4--6 and 13--19).
In the first group, with one exception (burst 21), oscillations appear only after the burst has reached 
its maximum flux, whereas in the second group the oscillations always appear right at the start of the burst.
Note that for burst 6, while we see the oscillations at the start of the burst, they appear later than in the other 
bursts with early oscillations. Burst 6 is also the burst which in its group has the strongest hint for radius 
expansion.

Previously, all oscillations observed in the tails of bursts appeared after the contraction phase of a radius 
expansion episode (Smith, Morgan \& Bradt 1997; Muno et al. 2000). Here we observe tail oscillations both in bursts 
with (e.g. burst 1) and without (e.g. burst 17) radius expansion. All bursts without radius expansion exhibited 
oscillations in the burst rise; the radius expansion bursts with oscillations (except burst 21) did not. So, in 
radius expansion bursts, as reported before, the oscillations appear after radius ``touchdown'' at the end of the 
radius expansion phase, but contrary to other sources (4U 1636--54, Strohmayer et al. 1998a; KS 1731--260, Muno et al. 
2000), in 4U 1728--34 there is a strong tendency for them to be entirely undetectable in the burst rise. Remember 
that the grey--scale power amplitudes in Figure \ref{fig:dynspec} refer to the 4 seconds around the point where they 
are plotted. 

Bursts 4, 5, and 17 show a "complete" 
frequency track, where in about 10 s the frequency rises by $\sim 3.5$ Hz before converging asymptotically to 
$\sim 364$ Hz. The initial rise in frequency seems to be steeper in burst 17 than in the other two. 
In all other bursts it seems that we only observe part of a full track such as seen in bursts 
4, 5, or 17. The oscillations in these bursts always show up in the time--frequency position where one 
would expect them based upon the shape of the frequency track in either bursts 4 and 5, or burst 17. 
We do not observe frequency decreases such as seen in one burst in 4U 1636--53 
(Miller 2000; Strohmayer 1999) and one burst in KS 1731--260 (Muno et al. 2000). 

\section{Discussion}

There have been several previous studies linking burst parameters to inferred mass 
accretion rate. Some studies used the persistent flux as an indicator for mass 
accretion rate (e.g. EXO 0748--676, Gottwald et al. 1986; 4U 1608--53, Murakami et 
al. 1980; van Paradijs, Penninx \& Lewin 1988). However, studies of the kilohertz QPOs 
and timing features at low frequency have shown that for the LMXBs the position in the 
color--color diagram is a better correlator to the timing properties and hence may be a better indicator 
for mass accretion rate than the 
persistent flux (Hasinger \& van der Klis (1989), for a review see van der Klis et al. 2000). For 4U 1636--52 
and 4U 1705--44 it was shown earlier that the correlation between burst parameters $kT_{\rm 0.1}$, 
$\tau$ and position of the source in the color--color diagram was strong, 
where the correlation with the persistent flux was lacking (van der Klis et al. 1990). In 4U 1728--34 we 
find no preference for the position in the color--color diagram over the persistent flux 
as a correlator with the burst spectral properties as there was with properties of the kilohertz QPOs (\mm 
\& van der Klis 1999) and with properties of several low frequency features (Di Salvo et al. 2000a). Perhaps 
this is due to the relatively large intrinsic scatter in the burst parameters. In this work we use the 
position in the color--color diagram, as in view of the results for other sources (see above) we expect this 
to be the more robust correlator.

We now compare our results with those recently obtained for KS 1731--260 
by Muno et al. (2000). 
In Table \ref{tbl:comp} we compare several properties of KS 1731--260 and 4U 1728--34.
KS 1731--260 showed only two distinct states, the island state 
and the banana branch with no data covering the transition 
between these two states. The bursts in the island state were longer ($\tau$ $\sim 15$ 
s) than those on the banana branch ($\sim 7$ s). The peak flux was slightly lower 
($\sim 4 \times 10^{-8}$ erg cm$^{-2}$ s$^{-1}$) in the island state than on the banana branch ($\sim 5 
\times 10^{-8}$ erg cm$^{-2}$ s$^{-1}$), and $E_{\rm b}$ was higher ($\sim 7 \times 
10^{-7}$ erg cm$^{-2}$) in the island state than on the banana branch ($\sim 4 \times 10^{-7}$ erg cm$^{-2}$). 
Radius expansion bursts were observed on the banana branch but not in the island state. 

In 4U 1728--34, contrary to the case of KS 1731--260, we observe a smooth transition between the island 
state and the banana branch. Di Salvo et al. (2000a) found that power spectra with 
rank numbers 1--11 are typical of the island state, 14 corresponds to the lower banana and 19 to the 
upper banana. We find that from island to banana $\tau$ changes only marginally, from $\sim 6$ s in the 
island state to $\sim 5$ s on the banana branch. Another source for which $\tau$ is about constant 
(and also short, with $\tau = 4.2 \pm 1.2$) over a wide range of $F_{\rm peak}$ and $E_{\rm b}$ (and we know 
that in 4U 1728--34 $F_{\rm peak}$ and $E_{\rm b}$ depend on rank number) is 4U 1735--44 (Lewin et al. 1980). 
However, most sources 
show the same behaviour as KS 1731--260, namely $\tau$ is several tens of seconds at low 
inferred mass accretion rate and about 5 s at high inferred mass accretion rate (e.g., EXO 0748--676, 
Gottwald et al. 1986; 4U 1636--53, van der Klis et al. 1990; 4U 1705--44, van der Klis 1989).
We note that a previous study of 4U 1728--34 by Basinska et al. (1984) did show a larger range in $\tau$ 
(5.1--12.3 s).

\begin{figure}
\figurenum{6}
\epsscale{0.7}
\plotone{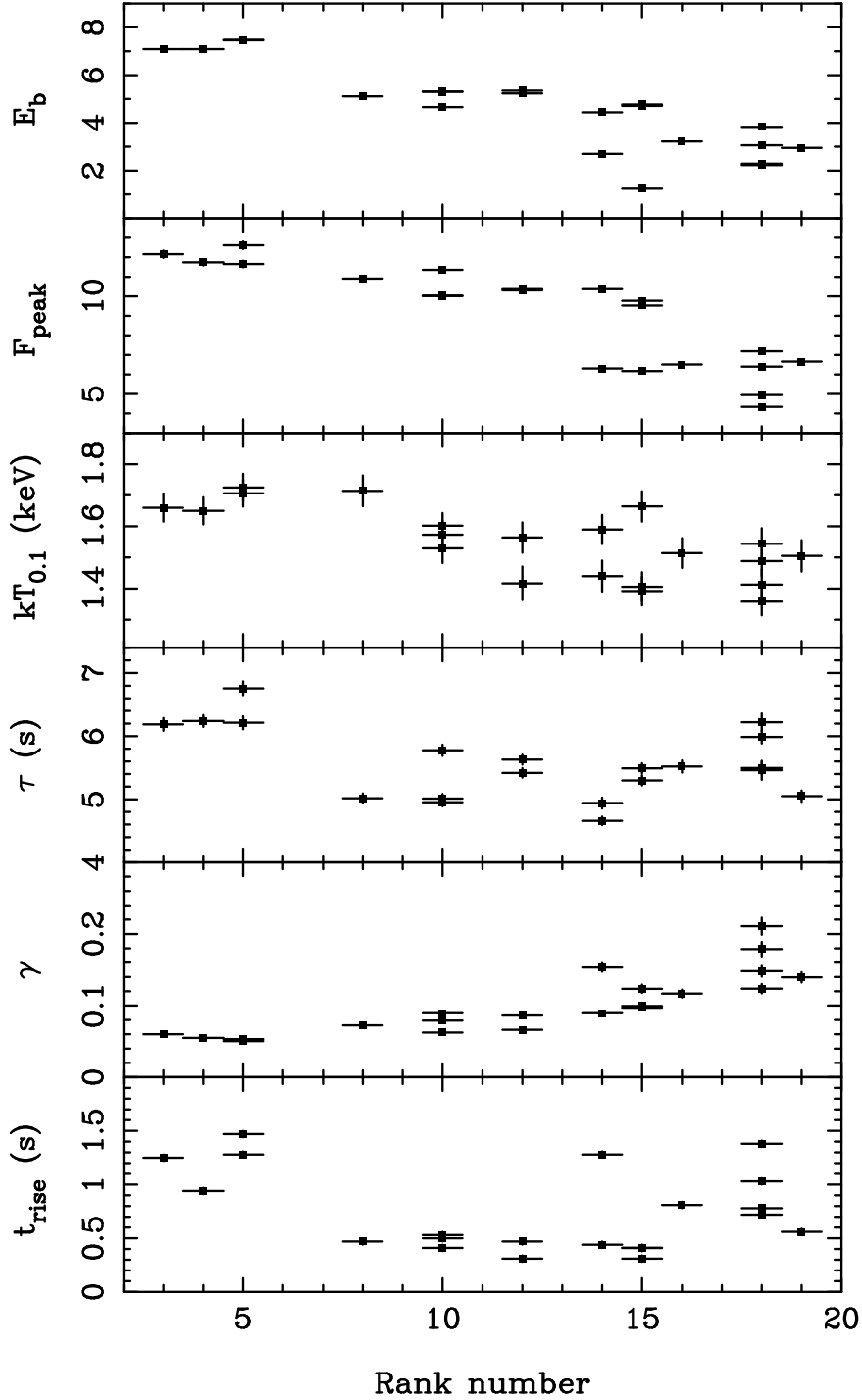}
\caption{Plotted versus rank number are the fluence, $E_{\rm b}$, the peak
flux, $F_{\rm peak}$, the blackbody temperature at 10 \% of the
Eddington flux, $kT_{\rm 0.1}$, $\tau$ ($=E_b/[F_{\rm peak}-F_{\rm pers}]$), $\gamma$
($=F_{\rm pers}/[F_{\rm peak}-F_{\rm pers}]$) and the rise time, $t_{\rm rise}$. The
short burst (number 19) with $\tau$ = 2.1 s was excluded from the $\tau$ plot; it was 
located at rank number 15. Two points at rank number 5 coincide in the top frame.}
\label{fig:rn_vs_sp}
\end{figure}

Another difference with KS 1731--260 is that in 4U 1728--34 the bursts in the island state 
show radius expansion (albeit with unusual radius variations), and have a higher $F_{\rm peak}$. 
As the source moves from the 
lower towards the upper banana the radius expansion episodes disappear and $F_{\rm peak}$ 
becomes smaller (see Figures \ref{fig:cd} and \ref{fig:rn_vs_sp}). The sources EXO 0748--676 
(Gottwald et al. 1986) and 4U 1705--44 (Langmeier et al. 1987) like KS 1731--260 but contrary to 
4U 1728--34 both show {\it higher} peak fluxes at higher inferred mass accretion rates. 
The fluence in 4U 1728--34 does show a similar behaviour to KS 1731--260: it changes gradually from about 
$7 \times 
10^{-7}$ erg cm$^{-2}$ in the island state to $\sim 3 \times 10^{-7}$ erg cm$^{-2}$ on the banana 
branch (see Fig. \ref{fig:rn_vs_sp}).
 
In KS 1731--260 oscillations were found only to occur on the banana branch and not in the island 
state (Muno et al. 2000). For 4U 1728--34 we also observe that burst oscillations appear at the higher 
inferred mass accretion states (see Fig. \ref{fig:cd}). Because we see a gradual transition between 
island and banana state, we can ascertain that oscillations are already detectable when the power 
spectrum of the persistent emission (Di Salvo et al. 2000a) still indicates an island state (see above). 

Cumming \& Bildsten (2000) explain the presence/absence of burst oscillations in KS 1731--260 
in terms of the presence/absence of a H--rich region. They show that if an oscillation is generated near 
the burning layers, it can escape coherently from a helium--rich region, but not from a H--rich region.
The bursts in KS 1731--260 show  
characteristics of pure He ignition (short durations, radius expansion and high peak flux) 
when the source is in the banana state, and of mixed H/He ignition when the source is in the 
island state; and oscillations occur only when there is pure He ignition. 
However, in 4U 1728--34 in the island state we observe that the burst oscillations disappear while the 
bursts still have all characteristics of pure He ignition. 

In KS 1731--260 the presence of burst oscillations and radius expansion were nearly perfectly correlated; 
all but one of the bursts with burst oscillations also had radius expansion, and none of the bursts without 
oscillations showed radius expansion. In contrast, in 4U 1728--34 radius expansion bursts occur below a 
certain inferred mass accretion rate level, burst oscillations above another level and in the middle both 
phenomena occur in the same bursts. It is striking that the bursts with radius expansion that show 
the unusual radius variations described in \S 2.1 are exactly the bursts from which the oscillations are absent.
We cannot from our data determine if unusual radius variations and absence of oscillations have a causal relation or 
are both just a characteristic of bursts occurring at low inferred mass accretion rate. Nevertheless, while there may 
not be a good correlation between the occurrence of radius expansion and burst oscillations, the way the radius 
expansion occurs may in fact be physically connected to the occurrence of oscillations. Perhaps, the unusual radius 
variations, which occur only in our most energetic bursts, are an indication that the magnetic fuel containment near 
the magnetic poles (a possible explanation for late burst oscillations, Miller 1999) is less efficient in stronger 
bursts.

So, it seems that both in KS 1731--260 and 
in 4U 1728--34 the timing as well as the spectral parameters of the bursts depend monotonically on the position of 
the source in the color--color diagram. Also, both in KS 1731--260 and 4U 1728--34 the presence/absence of the 
burst oscillations depends on the position in the color--color diagram in more or less the same way. 
However, the dependence of most of the spectral burst parameters on the position in the 
color--color diagram shows a clear discrepancy between the two sources, from which we are led to conclude 
that the correlations between spectral burst parameters and the occurrence of oscillations suggested from the KS 
1731--260 data is not universal. 

To summarize:
\begin{itemize} 
 
\item The burst oscillations become significantly weaker (and in 
fact undetectable) in the island state without the characteristics of the bursts switching 
from pure He to mixed H and He unstable burning, as occurs in KS 1731--260.

\item Radius expansion per se does not determine the presence or absence of burst 
oscillations. Both are more likely to occur at lower inferred mass accretion rates.

\item Contrary to what was noted in Muno et al. (2000) for other sources, we do observe oscillations 
in the tails of non radius expansion bursts. 

\item Among our bursts with radius expansion, the most energetic ones at the lowest inferred mass accretion 
rate, which show an unexpected spectral behaviour with a double--peaked radius curve do not show burst 
oscillations, while the others do.

\item Among our bursts with oscillations, the radius expansion bursts (with one exception) show no 
oscillations during the burst rise whereas non--radius expansion bursts do.

\end{itemize}
As the only two sources studied until now already show remarkable differences in their behavior, further 
studies of correlated spectral and timing behaviour of other type I X--ray bursters showing burst oscillations 
are necessary to come to a full understanding of just what determines whether, and if so, in which part of 
its profile, a burst oscillates.

\section{Acknowledgements}

This work was supported by NWO SPINOZA grant 08--0 to E.P.J. van den Heuvel, 
by the Netherlands Organization for Scientific
Research (NWO) under contract number 614--51--002, and by 
the Netherlands Research School for Astronomy (NOVA). 
This research has made use of data obtained through
the High Energy Astrophysics Science Archive Research Center Online Service, 
provided by the NASA/Goddard Space Flight Center. 
We would like to thank Tiziana di Salvo for providing the data of the 
color--color diagram. We would also like to thank the referee whose comments 
helped us to improve the paper.

{\it Note added in manuscript.}--Just before this paper was submitted we became aware of a paper in preparation 
by Franco (2000) on the burst oscillations in 4U 1728--34. Where between the two papers conclusions overlap, 
they are consistent. To avoid confusion, we note that our definition of a "complete frequency" track is more 
restrictive than "a burst with oscillations in the rise and the tail" used in Franco (2000). We call a frequency 
track "complete" only when the oscillations are present essentially throughout the entire burst ($\geq$ 14 s). 
We would like to thank Lucia Franco for her willingness to exchange manuscripts before publication.


\begin{deluxetable}{lccccc}
\tabletypesize{\small}
\tablewidth{0pt}
\tablecaption{Bursts from 4U 1728--34\label{tbl:obs}} 
\tablehead{ 
\colhead{Burst Number} & \colhead{Start Time} & 
\colhead{Intensity} & \colhead{Soft Color} & 
\colhead{Hard Color} & \colhead{Rank number}\\
\colhead{} & \colhead{(UTC)} & 
\colhead{(c/s)} & \colhead{} & 
\colhead{}  & \colhead{}
}

\startdata
1  & 15 Feb 1996 17:58:11 & 1627 & 2.82 & 0.52 & 10\\
2  & 15 Feb 1996 21:10:19 & 1800 & 2.79 & 0.50 & 12\\
3  & 16 Feb 1996 03:57:09 & 1933 & 2.83 & 0.48 & 14\\
4  & 16 Feb 1996 06:51:08 & 1980 & 2.82 & 0.47 & 15\\
5  & 16 Feb 1996 10:00:45 & 2043 & 2.83 & 0.48 & 15\\
6  & 16 Feb 1996 19:27:11 & 1940 & 2.78 & 0.48 & 14\\
7  & 18 Feb 1996 17:31:50 & 1414 & 2.79 & 0.53 & 10\\
8  & 18 Feb 1996 21:28:50 & 1317 & 2.85 & 0.56 &  6\\
9  & 22 Feb 1996 23:09:11 &  923 & 3.05 & 0.62 &  5\\
10 & 24 Feb 1996 05:46:23 &  891 & 3.01 & 0.62 &  5\\
11 & 24 Feb 1996 17:51:48 &  948 & 3.10 & 0.65 &  3\\
12 & 25 Feb 1996 23:17:06 &  998 & 3.13 & 0.62 &  4\\   
13 & 19 Sep 1997 12:32:55 & 1977 & 3.02 & 0.45 & 18\\  
14 & 20 Sep 1997 10:08:50 & 1954 & 3.03 & 0.44 & 18\\  
15 & 21 Sep 1997 15:45:28 & 2158 & 3.09 & 0.44 & 19\\ 
16 & 21 Sep 1997 18:11:04 & 2080 & 3.03 & 0.46 & 18\\ 
17 & 22 Sep 1997 06:42:51 & 1872 & 2.97 & 0.45 & 18\\  
18 & 26 Sep 1997 14:44:09 & 1620 & 2.75 & 0.46 & 16\\ 
19 & 26 Sep 1997 17:29:47 & 1554 & 2.74 & 0.47 & 15\\  
20 & 27 Sep 1997 11:19:06 & 1287 & 2.75 & 0.52 & 10\\  
21 & 27 Sep 1997 15:54:03 & 1365 & 2.78 & 0.50 & 12 
\enddata

\tablecomments{Listed for each burst are the burst number, 
the start time (the time the EDS was triggered, in Universal Time, 
Coordinated), the intensity (2.0--16.0 keV count rate in all five 
detectors of the PCA), soft color (3.5$-$6.4$/$2.0$-$3.5 keV count rate 
ratio), hard color (9.7$-$16.0$/$6.4$-$9.7 keV count rate ratio) 
and rank number, a measure for the position of the source in the color-color 
diagram prior to the burst (see text).}

\end{deluxetable}

\begin{deluxetable}{lcccccccc}
\tablecaption{Energy spectral properties of the bursts\label{tbl:spec}}
\tabletypesize{\footnotesize}
\tablewidth{0pt}
\tablehead{ 
\colhead{Burst} & \colhead{$F_{\rm peak}$} & 
\colhead{$E_{\rm b}$} & \colhead{$kT_{\rm 0.1}$} & 
\colhead{$\tau$} & \colhead{$\gamma$} &
\colhead{$\alpha$} & \colhead{$t_{\rm rise}$} &
\colhead{Radius}\\
\colhead{\#} & \colhead{($10^{-8}$ erg s$^{-1}$ cm$^{-2}$)} & 
\colhead{($10^{-7}$ erg cm$^{-2}$)} & \colhead{(keV)} & 
\colhead{(s)}  & \colhead{} &
\colhead{} & \colhead{(s)} &
\colhead{Expansion}
}

\startdata
1  &  10.0(1)  & 5.31(2) & 1.53(5) & 5.8(1) &  0.089(4) & $>33$  & 0.41  &	Y\\
2  &  10.4(1)  & 5.35(2) & 1.42(5) & 5.6(1) &  0.086(3) & $>32$  & 0.47  &	Y\\
3  &  6.3(1)   & 2.70(1) & 1.59(5) & 4.9(1) &  0.153(6) & $>84$  & 1.28  &	N\\
4  &  9.5(1)   & 4.77(2) & 1.39(5) & 5.5(1) &  0.097(4) & $>29$  & 0.41  &	N$^{*}$\\
5  &  9.8(1)   & 4.71(2) & 1.41(5) & 5.3(1) &  0.100(4) & $>23$  & 0.41  &	N$^{*}$\\
6  &  10.4(1)  & 4.44(2) & 1.44(5) & 4.7(1) &  0.089(3) & $>30$  & 0.44  &	N$^{*}$\\
7  &  10.0(1)  & 4.66(2) & 1.57(5) & 5.0(1) &  0.079(3) & $>2.7$ & 0.50  &	Y\\ 
8  &  10.9(2)  & 5.11(2) & 1.71(5) & 5.0(1) &  0.072(3) & $>39$  & 0.47  &	Y\\ 
9  &  12.6(2)  & 7.46(4) & 1.73(4) & 6.2(1) &  0.050(3) & $>22$  & 1.28  &	Y\\ 
10 &  11.7(2)  & 7.48(4) & 1.71(4) & 6.8(1) &  0.053(3) & $>15$  & 1.47  &	Y\\ 
11 &  12.2(2)  & 7.09(3) & 1.66(5) & 6.2(1) &  0.060(3) & $>0.5$ & 1.25  &	Y\\ 
12 &  11.7(2)  & 6.95(3) & 1.65(4) & 6.2(1) &  0.055(3) & $>26$  & 0.94  &	Y\\ 
13 &  7.2(1)   & 3.83(2) & 1.36(4) & 6.0(1) &  0.12(1)  & $>5$   & 1.03  &	N\\
14 &  4.9(1)   & 2.29(4) & 1.54(5) & 5.5(2) &  0.18(1)  & $>119$ & 1.38  &	N\\
15 &  6.7(1)   & 2.95(1) & 1.51(5) & 5.0(1) &  0.14(1)  & $>5$   & 0.56  &	N\\
16 &  6.4(1)   & 3.06(2) & 1.41(5) & 5.5(1) &  0.15(1)  & $>100$ & 0.72  &	N\\
17 &  4.3(1)   & 2.23(2) & 1.49(5) & 6.2(1) &  0.21(1)  & $>89$  & 0.78  &	N\\
18 &  6.5(1)   & 3.22(1) & 1.51(5) & 5.5(1) &  0.12(1)  & $>51$  & 0.81  &	N\\
19 &  6.2(1)   & 1.24(1) & 1.66(5) & 2.3(1) &  0.12(1)  & $>44$  & 0.31  &	N\\
20 &  11.4(1)  & 5.30(2) & 1.60(4) & 5.0(1) &  0.062(2) & $>20$  & 0.53  &	Y\\
21 &  10.3(1)  & 5.23(2) & 1.56(5) & 5.4(1) &  0.066(3) & $>10$  & 0.31  &	Y
\enddata

\tablecomments{Listed for each burst are the burst number, 
the maximum flux, $F_{\rm peak}$, the fluence, $E_{\rm b}$, the temperature at 10 \% of the Eddington flux,
$kT_{\rm 0.1}$, the standard burst parameters $\gamma$ (=$F_{\rm pers}/[F_{\rm peak}-F_{\rm pers}]$), $\tau$ 
($=E_b/[F_{\rm peak}-F_{\rm pers}]$) and a lower limit (because of South Atlantic Anomaly passages 
and earth occultations) for $\alpha$ ($=F_{\rm pers}/(E_b/\Delta t)$), the rise time, $t_{\rm rise}$ 
and whether or not the burst shows radius expansion. $^{*}$ Hint for radius expansion, see \S 2.1.}

\end{deluxetable}

\begin{deluxetable}{lcc}
\tabletypesize{\normalsize}
\tablewidth{0pt}
\tablecaption{Frequency shifts and endpoints in 4U 1728--34\label{tbl:freq_shft}} 
\tablehead{ 
\colhead{Burst \#} & \colhead{$\Delta \nu$ (Hz)} & \colhead{$\nu_{\rm max}$ (Hz)}
}

\startdata
4  & 3.5  & 364.5\\
5  & 3.75 & 364.25\\
14 & 3.0  & 364.0\\
17 & 3.5  & 364.0\\
\enddata

\tablecomments{Frequency shift ($\Delta \nu$) and frequency endpoint ($\nu_{\rm max}$) of the 
four strongest frequency tracks in 4U 1728--34. The uncertainty is equal to the bin size in frequency 
in Figure \ref{fig:dynspec} (0.25 Hz). Note that these listed values might be influenced by noise powers (on average one 
per plot, see text) that are accidentally situated near the beginning or end of the frequency tracks 
in Figure \ref{fig:dynspec}.}
\end{deluxetable}

\begin{deluxetable}{lccccccc}
\tabletypesize{\normalsize}
\tablewidth{0pt}
\tablecaption{A comparison between 4U 1728--34 and KS 1731--260\label{tbl:comp}} 
\tablehead{ 
\colhead{} & \multicolumn{3}{c}{KS 1731--260} & \colhead{~~~~~~~~~} &
\multicolumn{3}{c}{4U 1728--34}\\
\colhead{} & \colhead{island} & \colhead{~~~} &
\colhead{banana} & \colhead{} & \colhead{island} & \colhead{$\rightarrow$} &
\colhead{banana}
}

\startdata
$\tau$ (s)						& $\sim 15$	& & $\sim 7$	& & $\sim 6$	& & $\sim 5$\\
$F_{\rm peak}$ ($10^{-8}$ erg s$^{-1}$cm$^{-2}$)		& $\sim 4$	& & $\sim 5$	& & $\sim 12$	& & $\sim 6$\\
$E_{\rm b}$	(in $10^{-7}$ erg cm$^{-2}$)			& $\sim 7$	& & $\sim 4$	& & $\sim 7$	& & $\sim 3$\\
Burst oscillations					& N		& & Y		& & N		& & Y\\
Radius Expansion					& N		& & Y		& & Y$^{*}$	& & Y $\rightarrow$ N\\ 
H/He							& H+He		& & He		& & He		& & He\\
\enddata

\tablecomments{The changes in burst properties from the island state to the banana branch for both 
4U 1728--34 and KS 1731--260. Listed are $\tau$, $F_{\rm peak}$, 
$E_{\rm b}$, the occurrence of burst oscillations and the occurrence of radius 
expansion. Also is shown whether the bursts show the characteristics of unstable pure He or mixed H/He 
burning. The arrow between island and banana for 4U 1728--34 indicates a smooth transition between the 
island state and the banana branch. The arrow between Y and N for radius expansion in the banana state 
of 4U 1728--34 indicates the presence of radius expansion in the lower banana and its absence in the 
upper banana. $^{*}$ Unusual radius variations, see \S 2.1.}

\end{deluxetable}

\onecolumn

\clearpage

\begin{figure}
\figurenum{2}
\epsscale{0.5}
\begin{tabular}{ccc}
\plotone{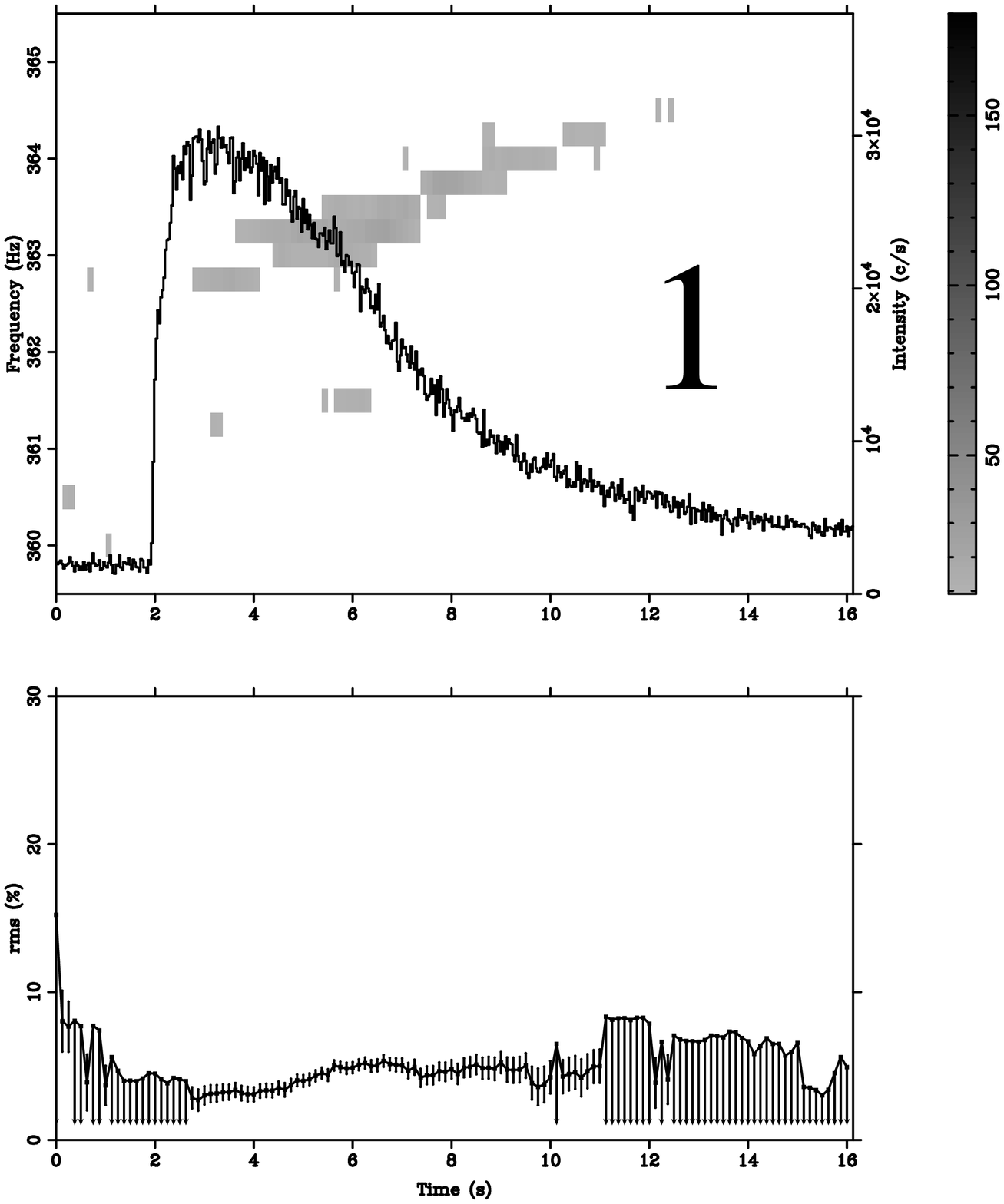} & \plotone{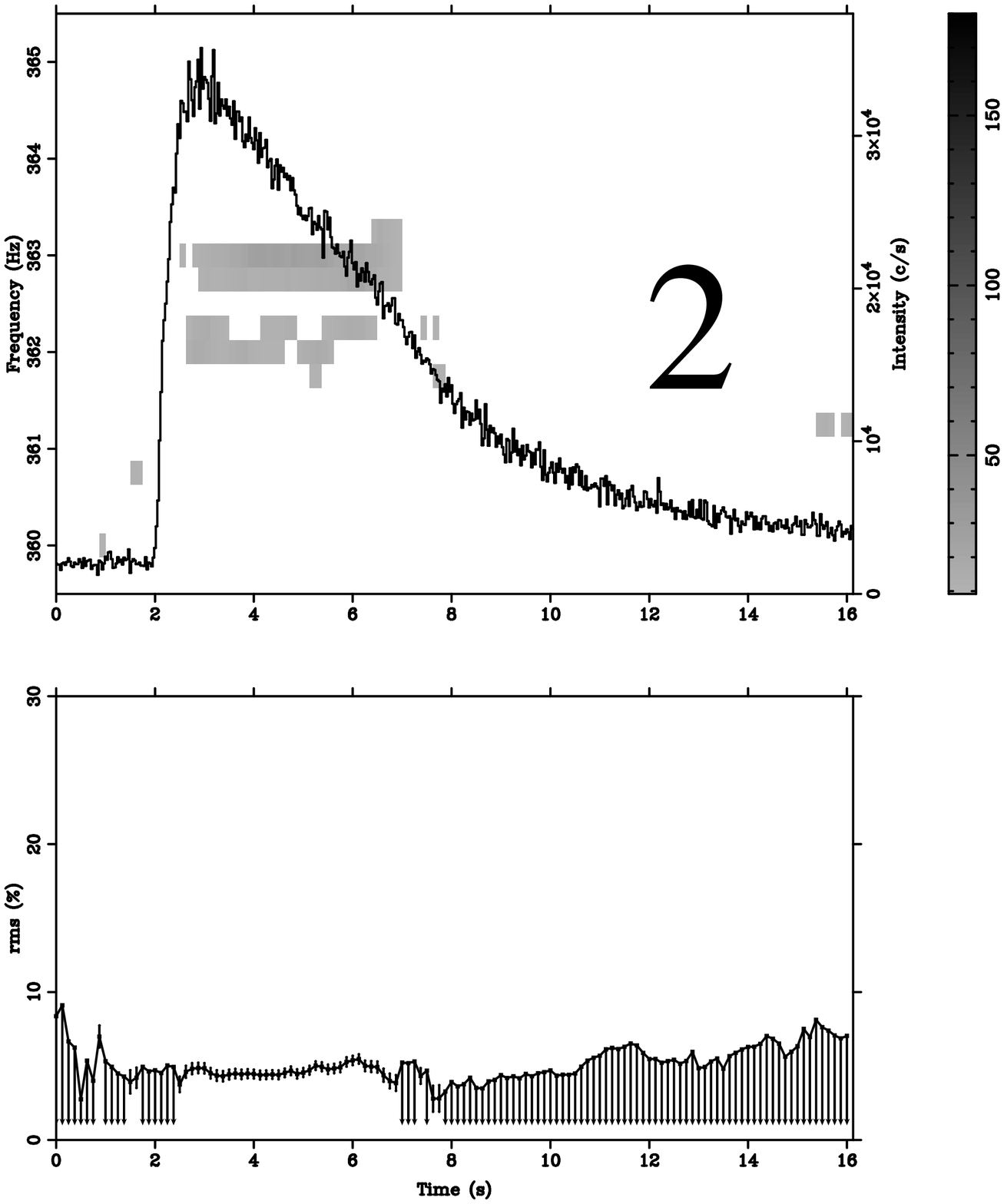} & \\
\plotone{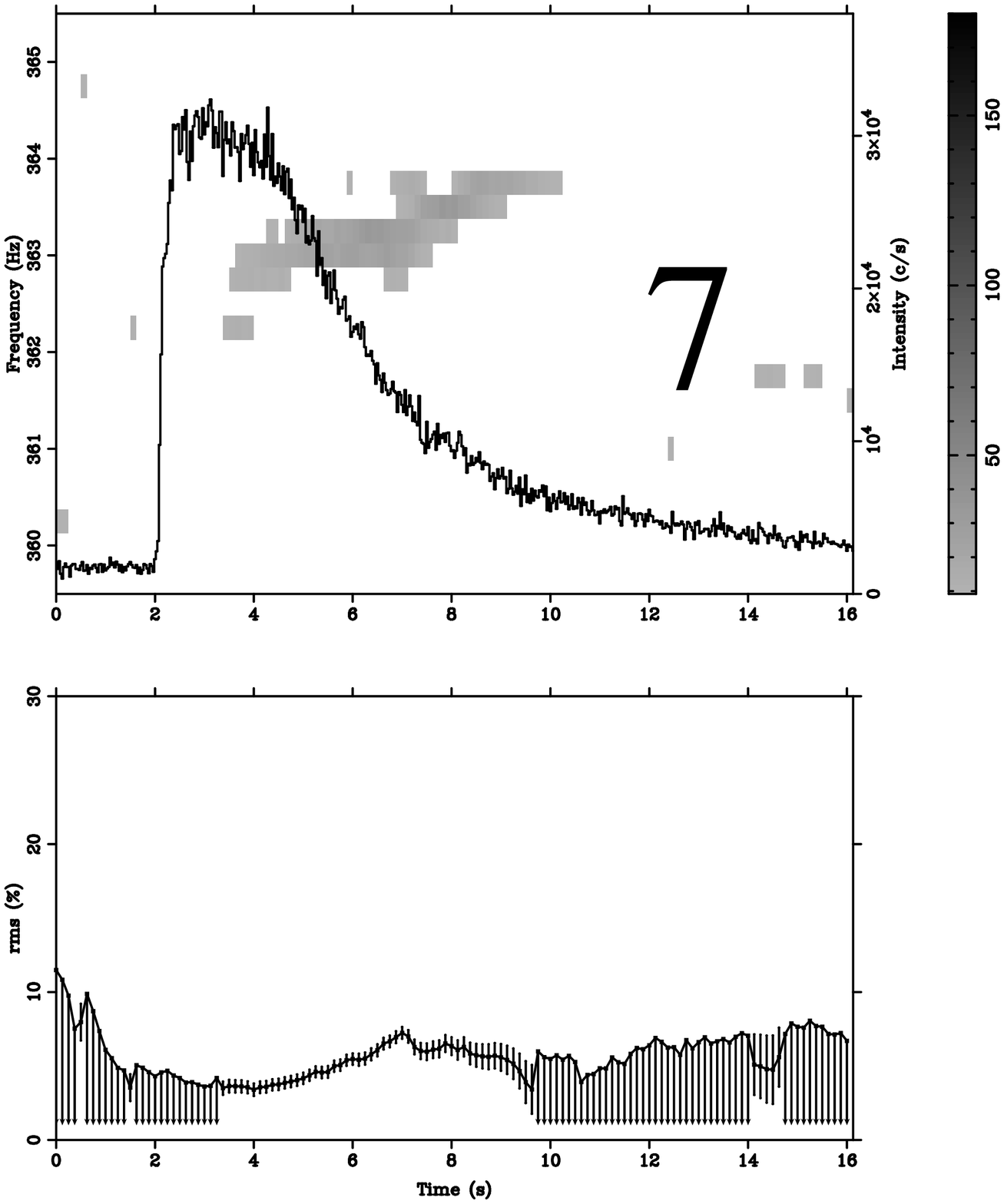} & \plotone{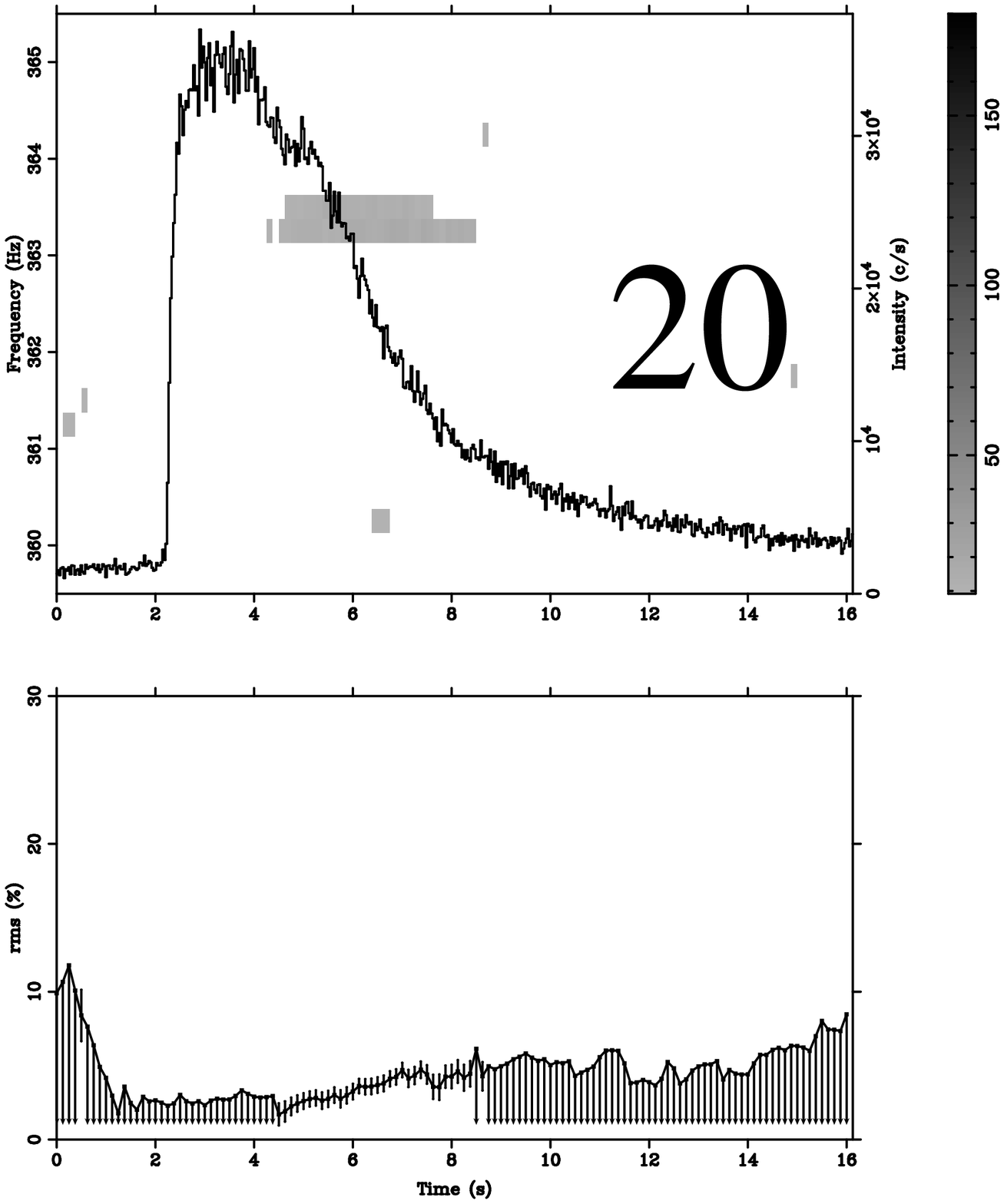} & \\
\end{tabular}
\caption{Atlas of burst oscillations in 4U 1728--34. Shown are time overlapping dynamic power 
spectra and associated PCA lightcurves (top frames), and fractional rms amplitudes of the 
oscillations (bottom frames). Arrows indicate 97.5 \% confidence upper limits. The first five 
bursts plotted (1, 2, 7, 20, 21) are the radius expansion bursts. Burst sequential numbers 
are indicated.}
\label{fig:dynspec}
\end{figure}
\clearpage

\begin{figure}
\figurenum{2}
\epsscale{0.5}
\begin{tabular}{ccc}
\plotone{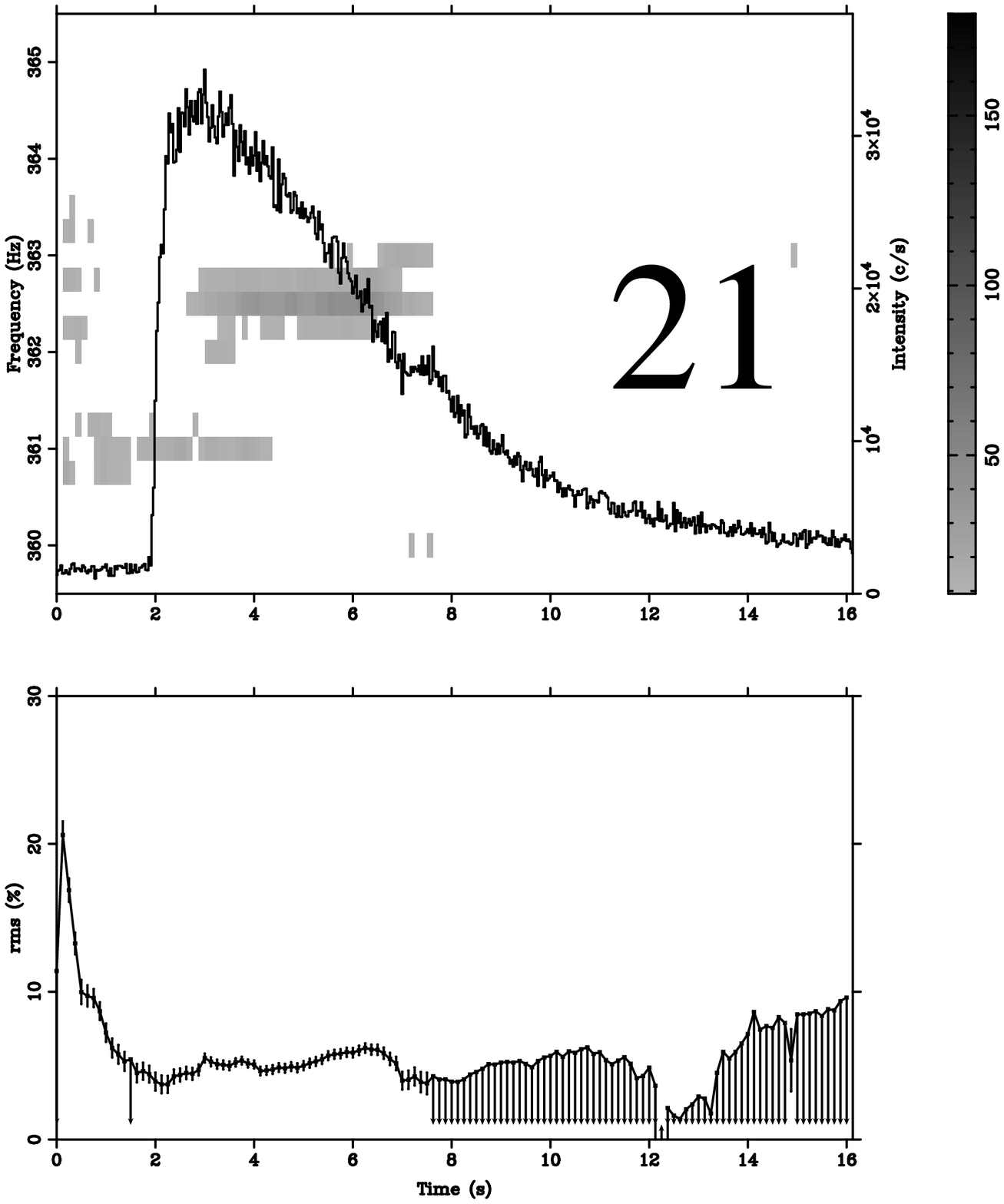} & \plotone{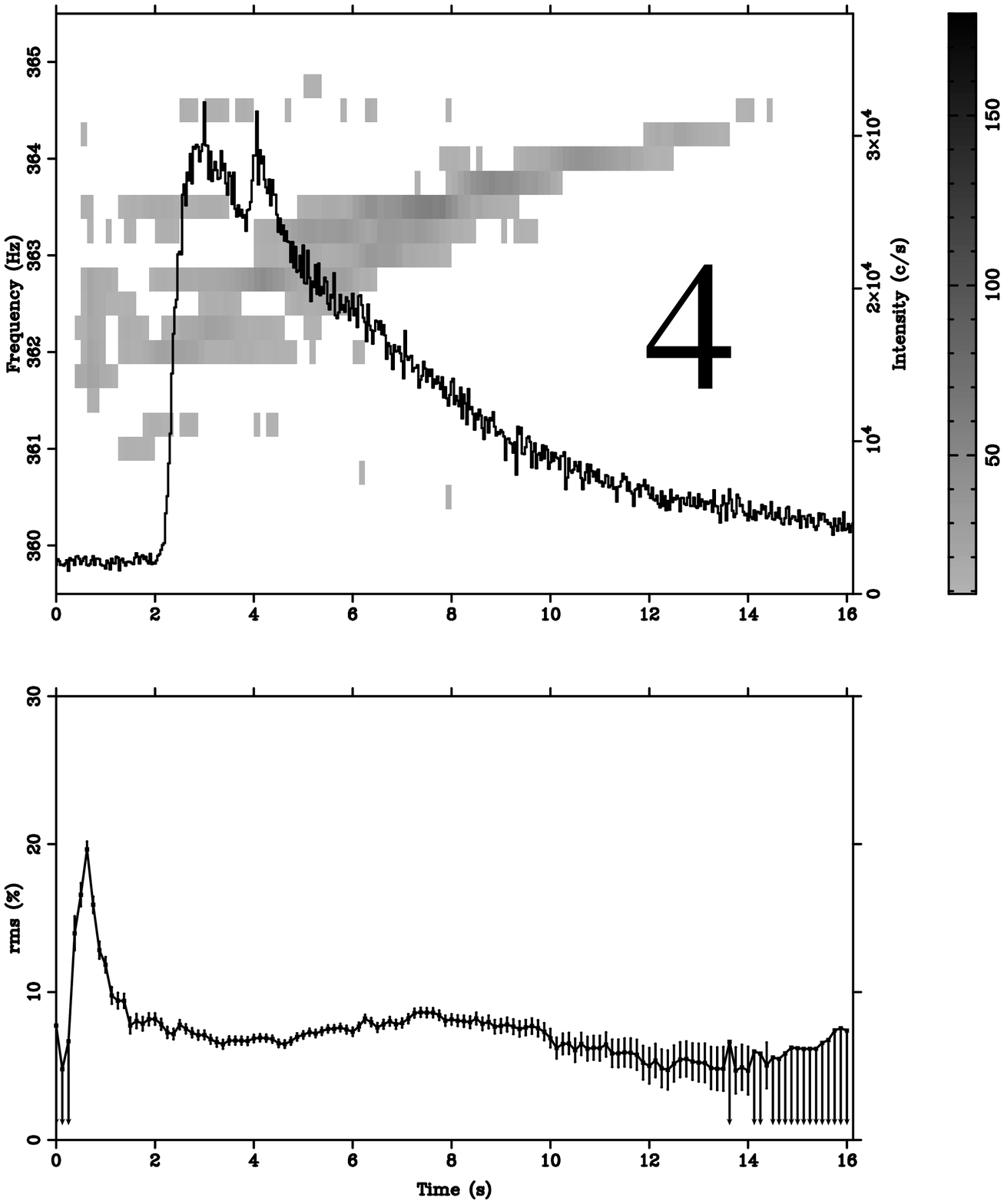} & \\
\plotone{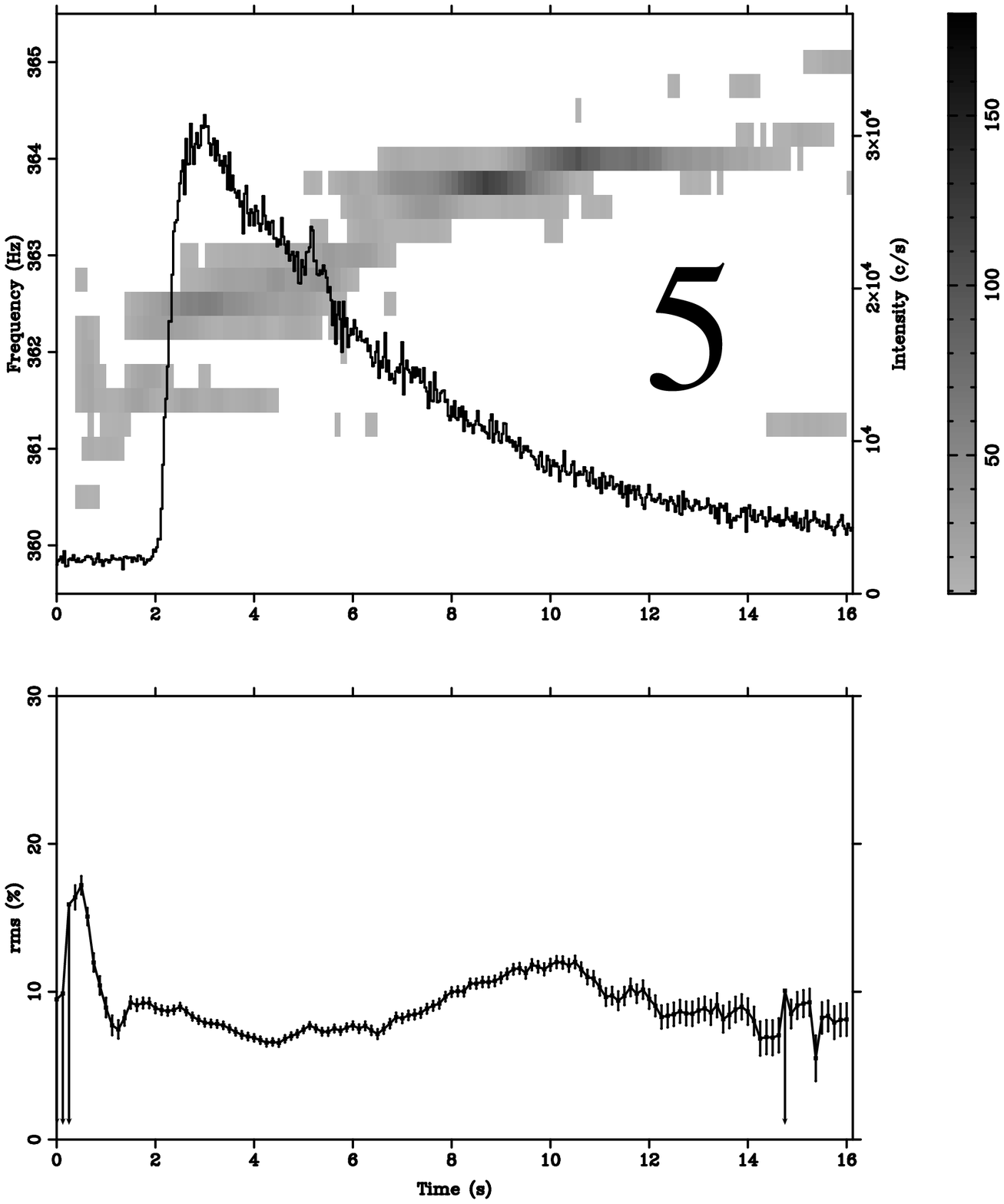} & \plotone{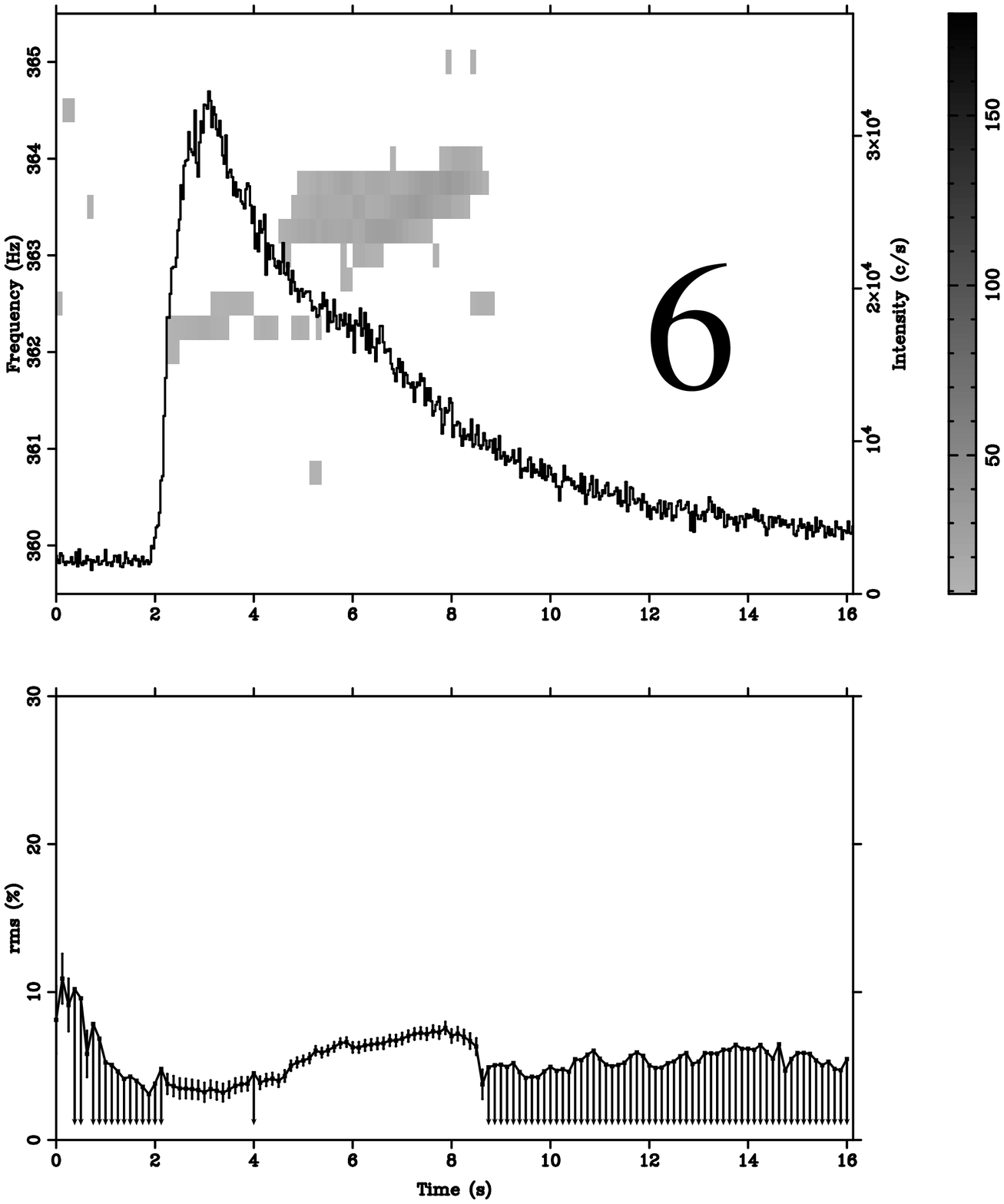} & \\
\end{tabular}
\caption{Continued}
\end{figure}
\clearpage

\begin{figure}
\figurenum{2}
\epsscale{0.5}
\begin{tabular}{ccc}
\plotone{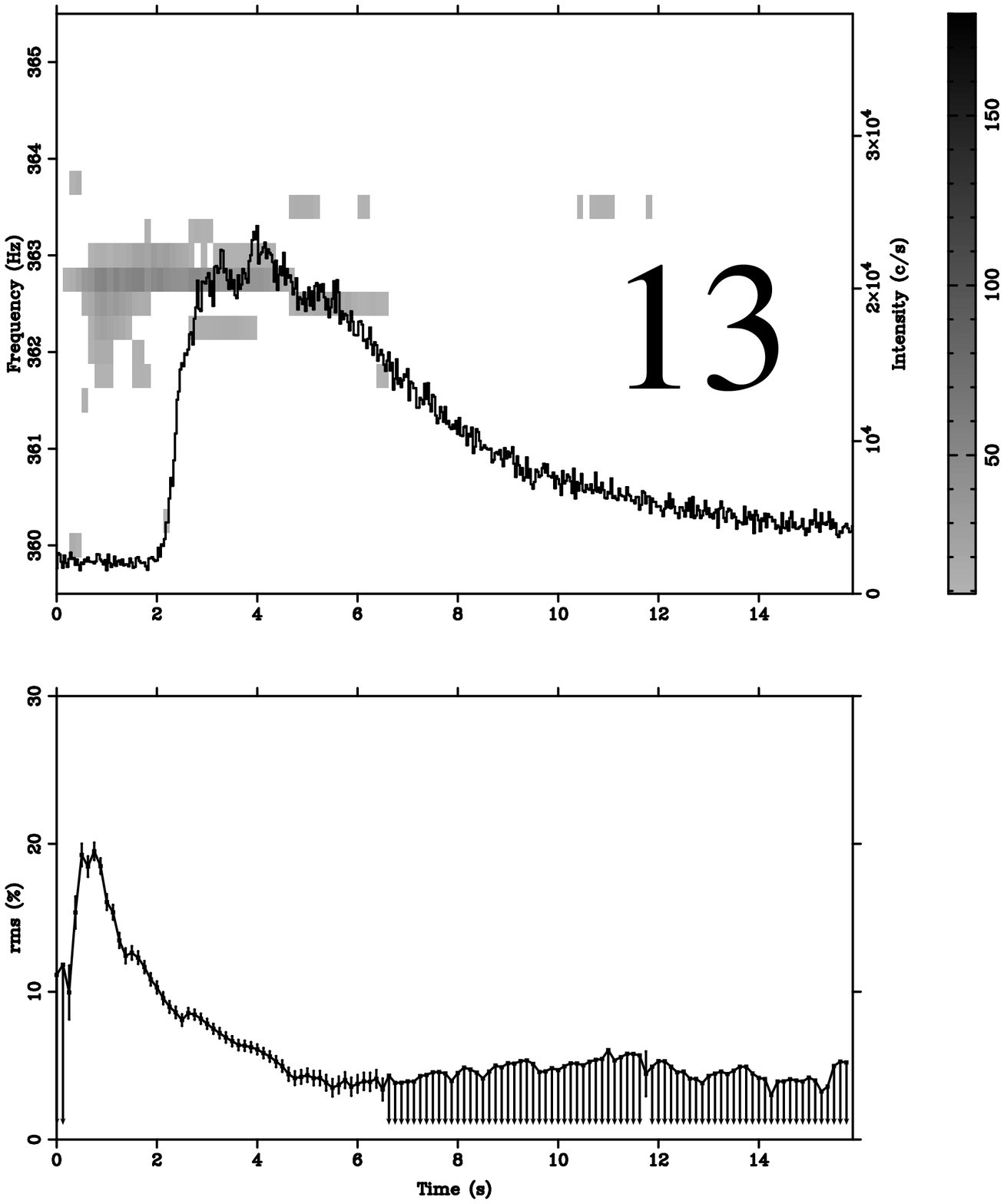} & \plotone{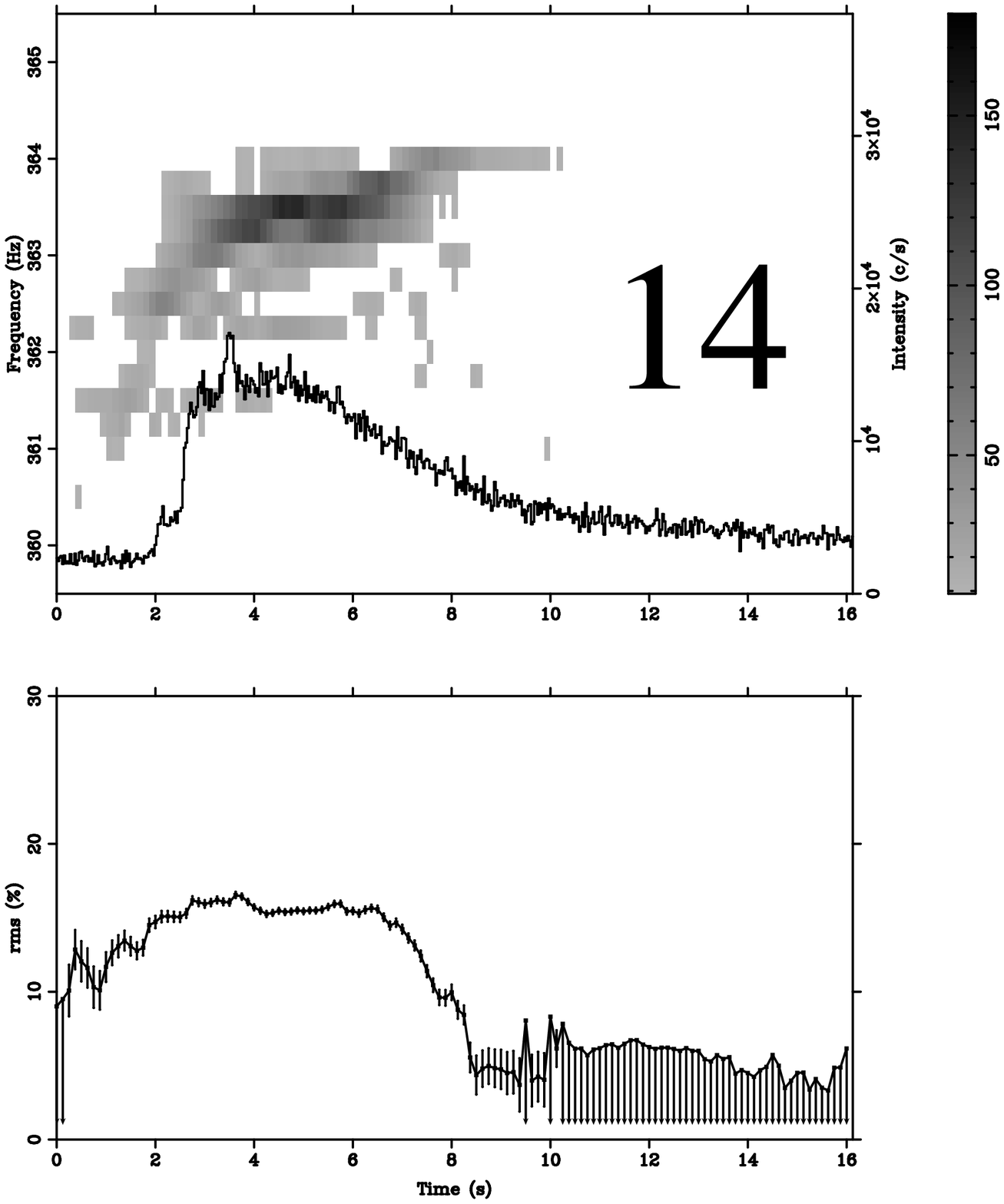} & \\
\plotone{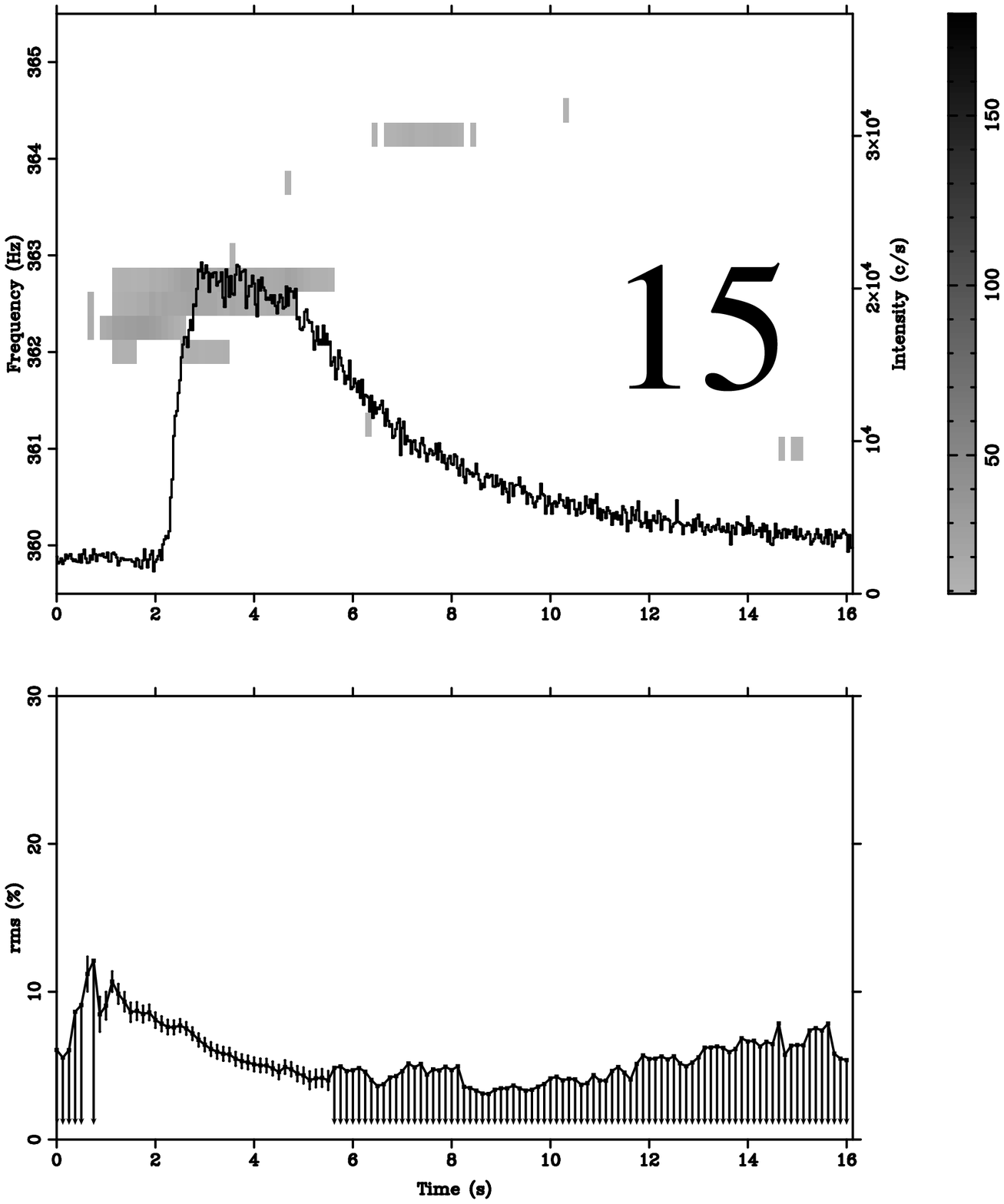} & \plotone{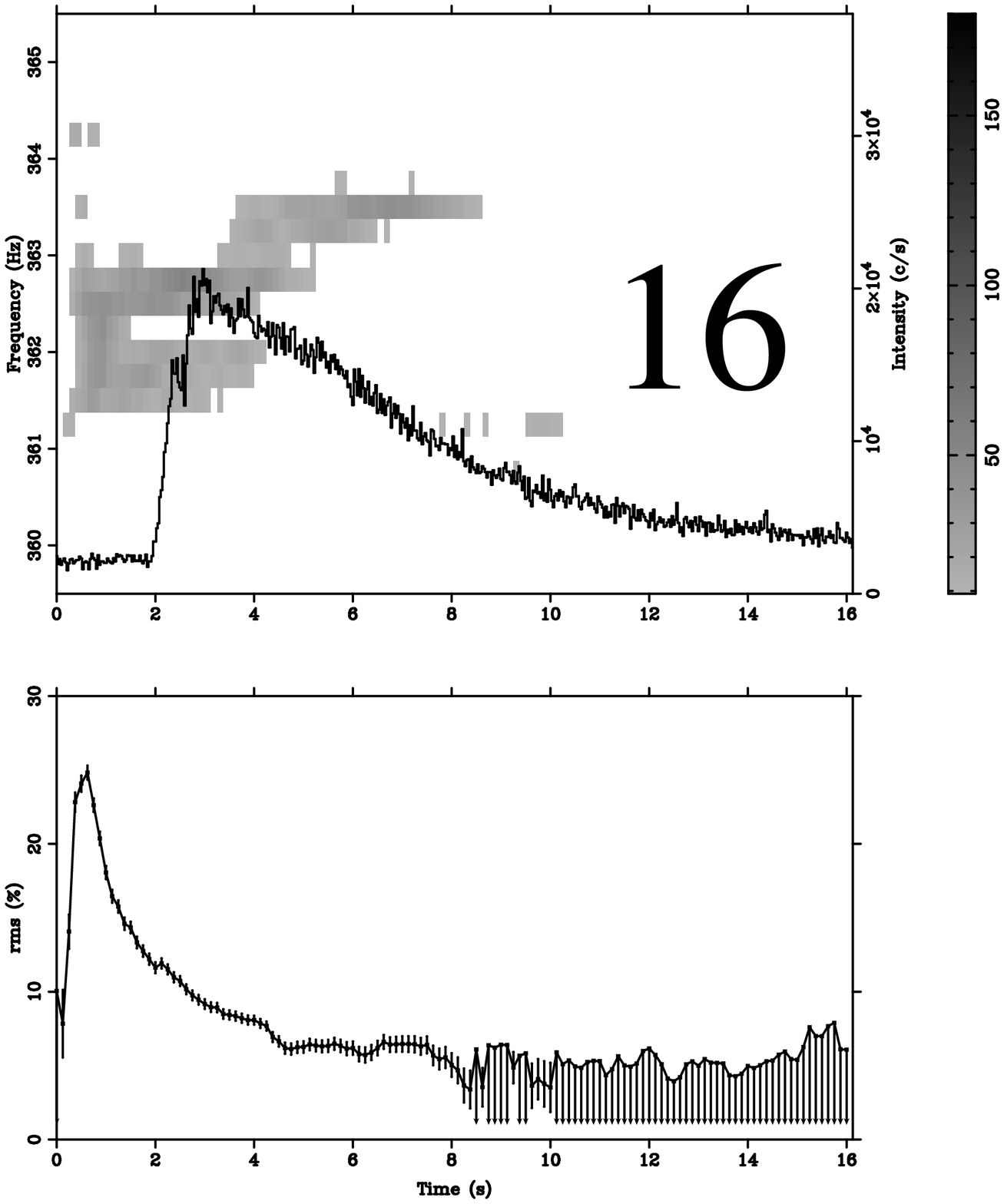} & \\
\end{tabular}
\caption{Continued}
\end{figure}
\clearpage

\begin{figure}
\figurenum{2}
\epsscale{0.5}
\begin{tabular}{ccc}
\plotone{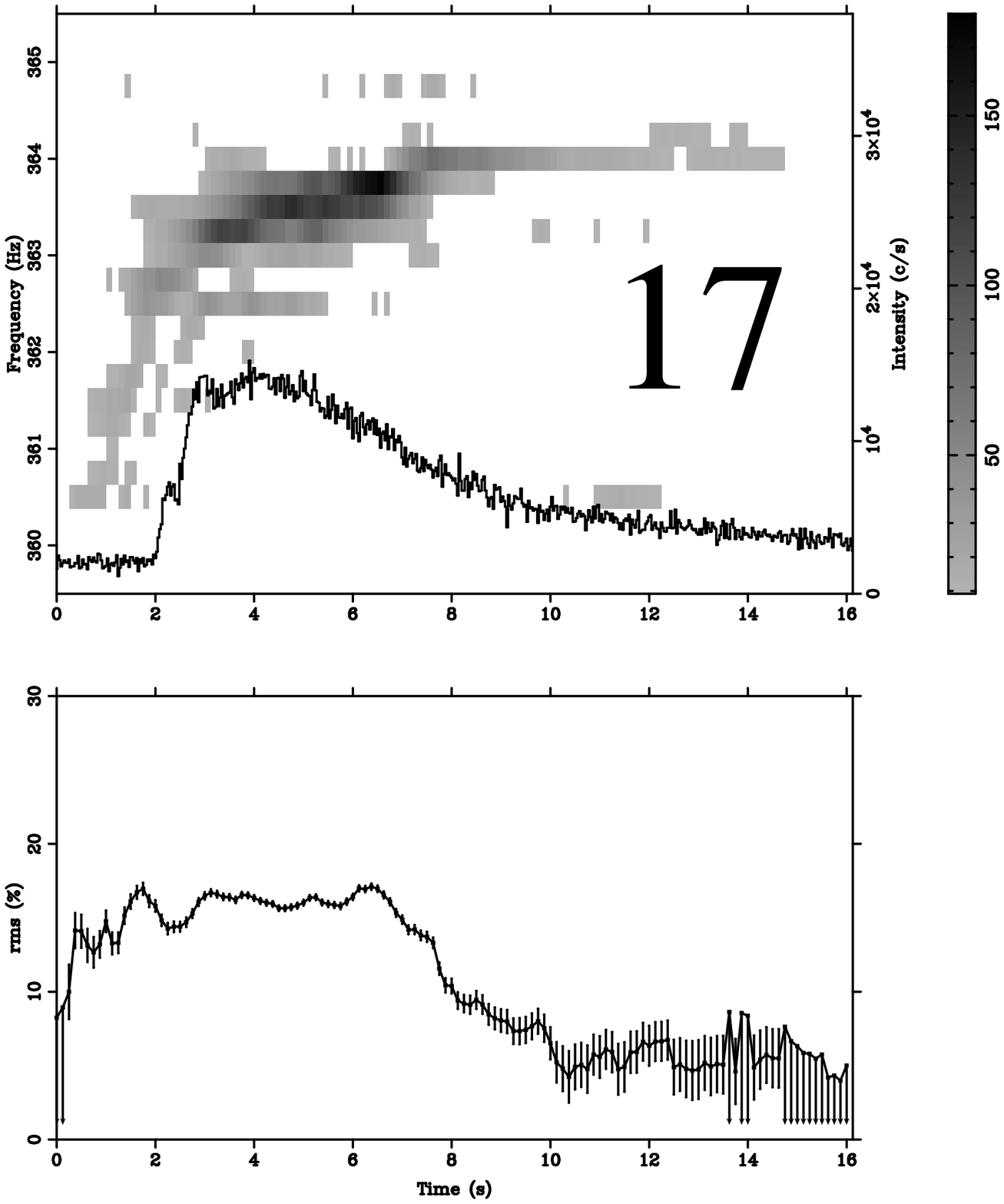} & \plotone{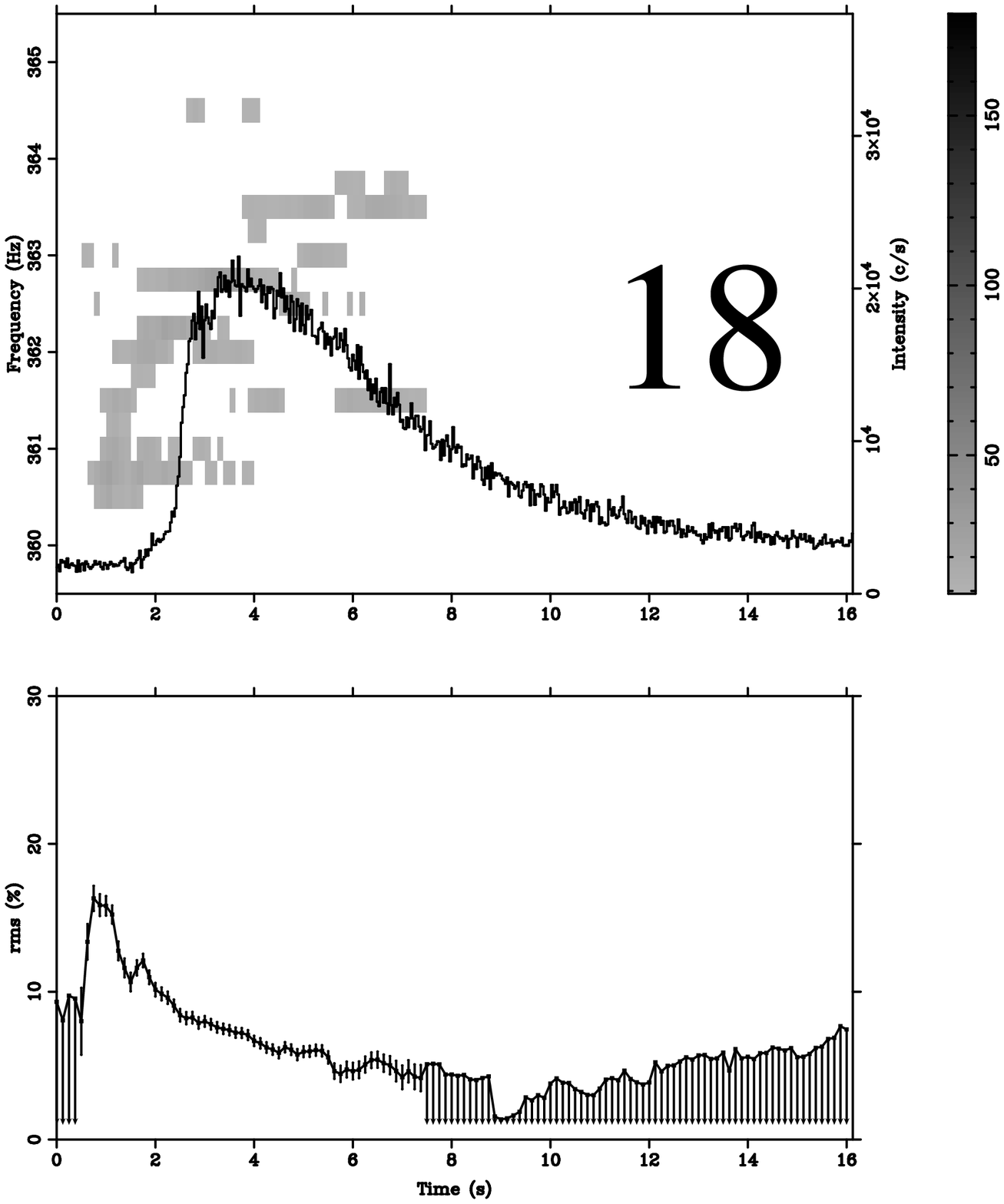} & \\
\plotone{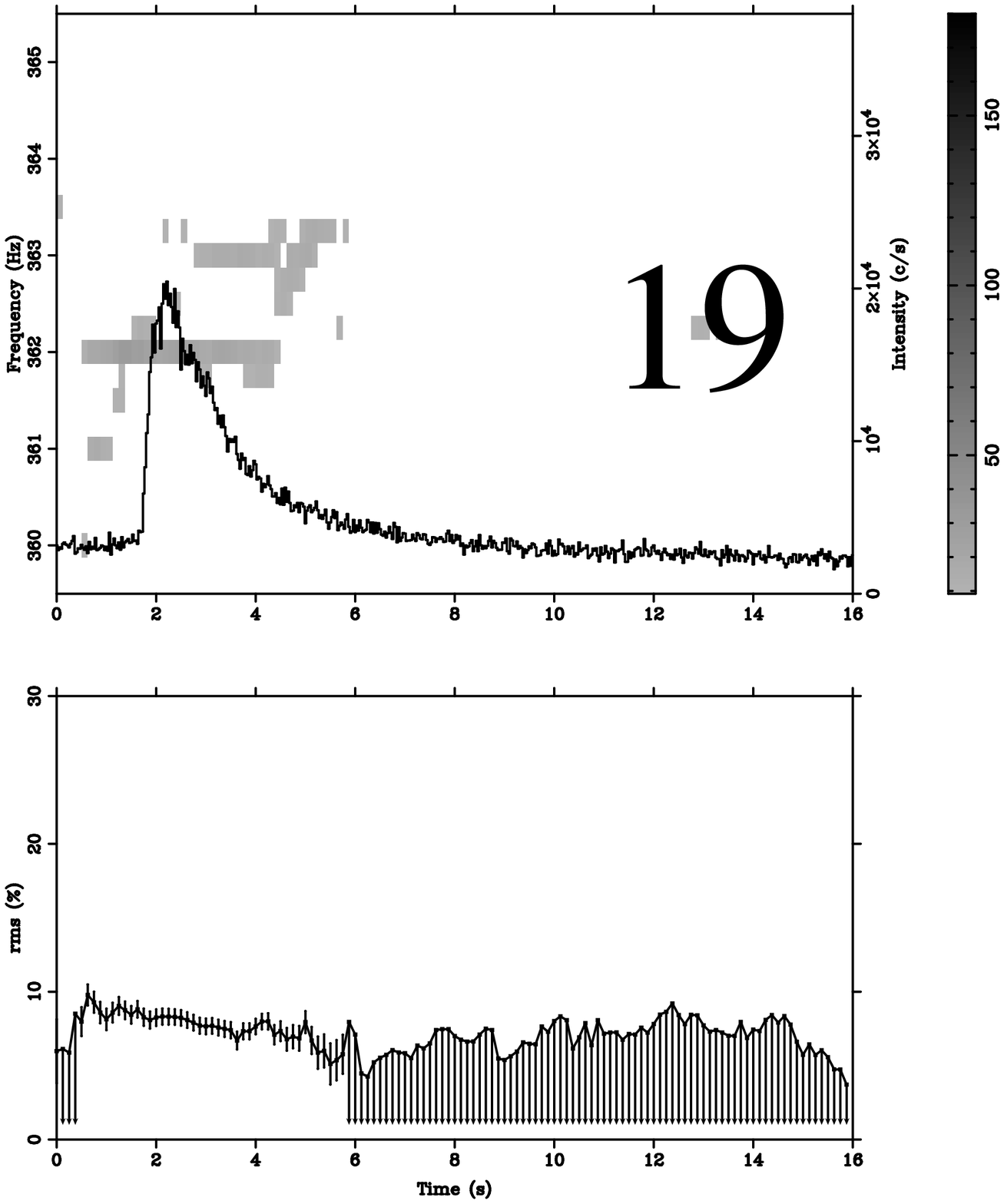} & & \\
\end{tabular}
\caption{Continued}
\end{figure}
\clearpage

\clearpage

\clearpage

\clearpage

\clearpage

\clearpage

\end{document}